\documentclass[aps,prb,twocolumn,superscriptaddress,
groupedaddress,10pt]{revtex4}
\usepackage{amsmath}
\usepackage{amssymb}
\usepackage{graphicx}
\usepackage{epsfig}
\usepackage{dcolumn}
\usepackage{bm}
\usepackage{blindtext}
\usepackage{natbib}
\usepackage{siunitx} 
\usepackage{multirow}
\usepackage{bbm}

\usepackage[usenames,dvipsnames]{xcolor}
\usepackage{amsthm}
\usepackage{booktabs}
\usepackage{hyperref}
\usepackage{tikz}
\usetikzlibrary{calc}
\usepackage{mathtools}

\def\be{\begin{equation}} \def\ee{\end{equation}}
\def\bea{\begin{eqnarray}} \def\eea{\end{eqnarray}}

\def\nn{\nonumber}

\newcommand{\ket}[1]{| #1 \rangle}
\newcommand{\bra}[1]{\langle #1 |}

\begin{document}
\title{Left-left-right-right magnetic order in spin-1/2 Kitaev-Heisenberg chain 
}

\author{Wang Yang}
\affiliation{School of Physics, Nankai University, Tianjin 300071, China}

\author{Chao Xu}
\affiliation{Institute for Advanced Study, Tsinghua University, Beijing 100084, China}
\affiliation{Kavli Institute for Theoretical Sciences, University of Chinese Academy of Sciences, Beijing 100190, China}

\author{Shicheng Ma}
\affiliation{School of Physics, Nankai University, Tianjin 300071, China}

\author{Alberto Nocera}
\affiliation{Department of Physics and Astronomy and Stewart Blusson Quantum Matter Institute, 
University of British Columbia, Vancouver, B.C., Canada, V6T 1Z1}

\author{Ian Affleck}
\affiliation{Department of Physics and Astronomy and Stewart Blusson Quantum Matter Institute, University of British Columbia, Vancouver, B.C., Canada, V6T 1Z1}

\begin{abstract}

In this work, we perform a combination of analytical and numerical studies on the phase diagram of the spin-1/2 Kitaev-Heisenberg chain in the region of negative Kitaev and positive Heisenberg couplings.
Apart from the antiferromagnetic phase, we find a magnetically ordered phase with left-left-right-right order and a gapless phase with central charge value $c=1$,
resolving the existing contradictory results in literature regarding  this parameter region.
In particular, the origin of the left-left-right-right order is clarified based on a perturbative Luttinger liquid analysis. 
The left-left-right-right phase is further shown to persist in the Kitaev-Heisenberg-Gamma model when a small nonzero Gamma interaction is introduced.  
Using a coupled-chain method, we also demonstrate  that the one-dimensional (1D) left-left-right-right order gives a quasi-1D explanation for  the 2D stripy order of the same model on the honeycomb lattice.

\end{abstract}

\maketitle

\section{Introduction}
\label{sec:intro}

One-dimensional (1D) generalized Kitaev spin models can be constructed from their two-dimensional (2D) counterparts \cite{Kitaev2006,Nayak2008,Jackeli2009,Chaloupka2010,Rau2014,Kimchi2014,Wang2017} by selecting one row out of the honeycomb lattice,
where the word ``generalized" is used to emphasize the fact that additional interactions beyond the Kitaev interaction are also included in the model in order to describe real Kitaev materials \cite{Witczak-Krempa2014,Rau2016,Winter2017,Hermanns2018}.  
Recently, motivated by the potential of the 1D studies in providing hints for the 2D Kitaev physics,
there have been surging research efforts in studying 1D generalized Kitaev models \cite{Sela2014,Gruenewald2017,Agrapidis2018,Agrapidis2019,Catuneanu2019,You2020,You2020b,Yang2019,Yang2020,Yang2020b,Yang2021b,Luo2021,Luo2021b,Sorensen2023,Sorensen2021,Yang2022a,Yang2022_2,Yang2022,Yang2022d,Yang2022e,Yang2022f,Zhao2022,Laurell2022}. 
On the other hand, in addition to being helpful for understanding 2D physics, 
1D generalized Kitaev models contain intriguing strongly correlated physics on their own, 
and it has been established that these 1D models have complicated nonsymmorphic symmetry group structures \cite{Yang2022a}, 
leading to exotic properties including emergent SU(2)$_1$ conformal invariance \cite{Yang2019,Yang2022d,Yang2022e},
nonsymmorphic bosonization  \cite{Yang2022_2,Yang2022}, 
nonlocal string order parameters \cite{Catuneanu2019,Luo2021},
solitons \cite{Sorensen2023},
and magnetic phases  breaking nonsymmorphic symmetries \cite{Yang2020}. 

The Kitaev-Heisenberg model \cite{Jackeli2009,Chaloupka2010} is an intensively studied prototypical type of generalized Kitaev models.  
The phase diagram of the 1D spin-1/2 Kitaev-Heisenberg model has been investigated in various works \cite{Sela2014,Agrapidis2018,Catuneanu2019,Yang2020}.
However, there are contradictory statements on the physics of  this model in the vicinity of the ferromagnetic (FM) Kitaev point. 
There is a special point in the FM Kitaev region located at $K=-2J<0$, which has a hidden SU(2) symmetry  \cite{Chaloupka2013,Chaloupka2015}.
The problem concerns with the phase diagram between the $K=-2J<0$ point and the FM Kitaev point: 
While Ref. \onlinecite{Sela2014} claims that the region is a $c=1$ Luttinger liquid phase where $c$ represents the central charge,
it was found in Ref. \onlinecite{Agrapidis2018,Yang2020} that the system is magnetically ordered.
In addition to the discrepancies in the literature on the nature of the phase diagram of the model,
there is no theoretical understanding on the origin of either the Luttinger liquid or  the magnetic order in the region. 

\begin{figure*}[ht]
\begin{center}
\includegraphics[width=12cm]{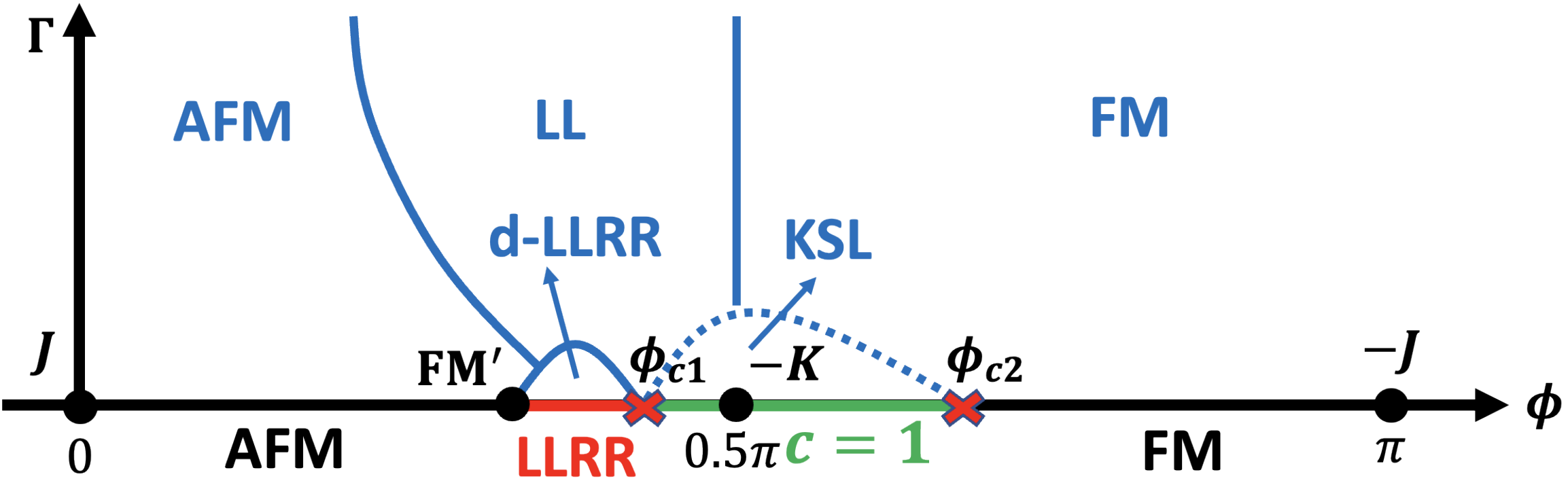}
\caption{Schematic plot of the phase diagram of the 1D spin-1/2  Kitaev-Heisenberg-Gamma model, where the horizontal axis (i.e., the $\Gamma=0$ line) corresponds to the Kitaev-Heisenberg model. 
The phase diagram in the region $\phi\in(0,\pi)$ close to the $\Gamma=0$ line is shown,
where the coordinate $\phi$ of the horizontal axis is defined in Eq. (\ref{eq:parametrization}) as $K=-\sin(\phi)$, $J=\cos(\phi)$. 
The $J$, FM$^\prime$, $-K$, $J$ points on the horizontal axis have $\phi$-values $0$, $\arctan(2)$, $\pi/2$, $\pi$, respectively,
where FM$^\prime$ is the hidden SU(2) symmetric point satisfying $K=-2J<0$.
The vertical axis is $\Gamma$, and because of the equivalence in Eq. (\ref{eq:equiv_Gamma}), only the $\Gamma>0$ region is shown. 
In the $\Gamma\neq 0$ region, ``d-LLRR" is ``distorted-left-left-right-right" for short;
``LL" is ``Luttinger liquid" for short; 
and ``KSL" is ``Kitaev spin liquid" for short, in which region the nature of the physics remains to be explored.
} \label{fig:phase}
\end{center}
\end{figure*}

Motivated by the above questions, in this work, 
we perform a combination of analytical and numerical studies in the region $\phi\in(\arctan{2},\pi/2)$,
in which $K=-\sin(\phi)$ and  $J=\cos(\phi)$ are the Kitaev and Heisenberg interactions, respectively. 
Notice that the hidden SU(2) symmetric FM point is located at $\phi=\arctan(2)$.
By a combination of perturbative Luttinger liquid analysis and density matrix renormalization group (DMRG) numerical simulations,  
we find that there is a magnetically ordered phase close to the $K=-2J$ point shown by ``LLRR" on the horizontal axis of the phase diagram in Fig. \ref{fig:phase},
where the spin ordering exhibits a collinear left-left-right-right (LLRR) pattern,
which is consistent with the finding in Ref. \onlinecite{Agrapidis2018} (named as staggered-$xy$ order there) and
different from the spiral phase identified  in Ref. \onlinecite{Yang2020}.
The key spirit of the strategy is straightforward to understand.
After a unitary transformation called four-sublattice rotation, 
the Hamiltonian can be decomposed into two parts, 
one part the easy-plane XXZ model, and the other part a four-site periodic term.
Since the Luttinger parameter diverges when the system approaches the $K=-2J$ point,
the system is very sensitive to perturbations, and as a result,  driven into an ordered state by the four-site periodic term.  
The predictions are supported by our DMRG \cite{White1992,White1993,Schollwock2011} simulations.

\begin{figure}[h]
\begin{center}
\includegraphics[width=8cm]{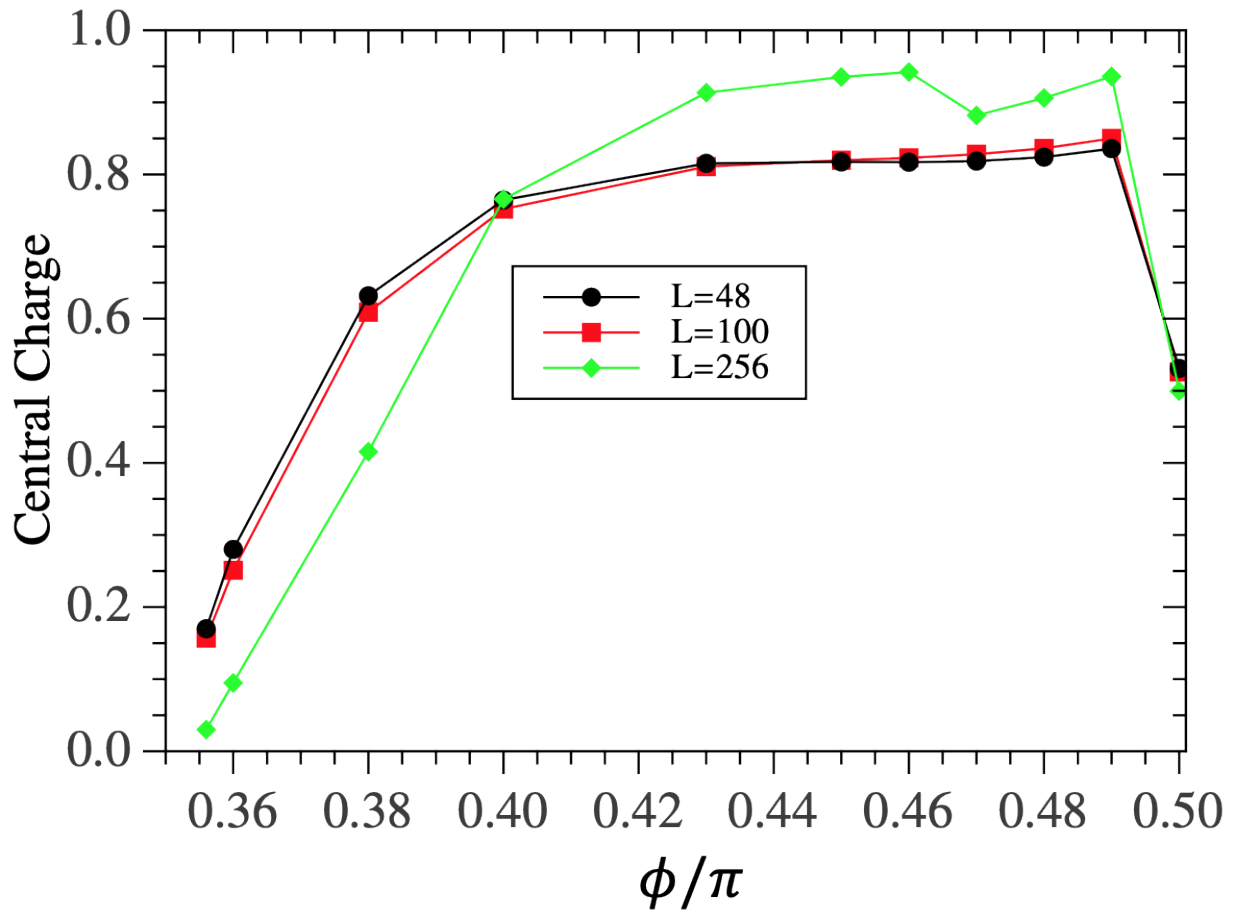}
\caption{Central charge values of the spin-1/2 Kitaev-Heisenberg chain as a function of $\phi$.
DMRG numerics are performed for $L=48,100,256$ sites using open boundary conditions. 
The bond dimension $m$ and truncation  error $\epsilon$ in DMRG calculations are taken as $m=1000$, $\epsilon=10^{-9}$.
} \label{fig:central_charge}
\end{center}
\end{figure}

Furthermore, as shown in Fig. \ref{fig:central_charge},  our DMRG numerics indicate that there exists   a phase transition point $\phi_{c1}\approx 0.4\pi$ which separates the LLRR phase from a $c=1$ gapless phase. 
Therefore, unlike the claims in the literatures \cite{Sela2014,Agrapidis2018,Yang2020},
the system have both ordered and gapless phases in the region $\phi\in(\arctan(2),\pi/2)$.
Since the $c=1$ phase is close to the FM Kitaev point which has exponentially large ground state degeneracy, 
the numerics in the gapless phase are very difficult. 
An analytical understanding of the gapless phase remains unclear and will be left for future investigations. 

We make two extensions of the LLRR phase in the 1D spin-1/2 Kitaev-Heisenberg model.
First, by adding a small nonzero Gamma term, we find that the LLRR phase extends to a finite region in the phase diagram of the spin-1/2 Kitaev-Heisenberg-Gamma chain,
having a non-collinear magnetic ordering pattern, 
which can be viewed as a slightly distorted version of the LLRR ordering, 
denoted as ``d-LLRR" in Fig. \ref{fig:phase}, where ``d" is ``distorted" for short.
Second, by weakly coupling an infinite number of chains on the honeycomb lattice, 
we recover the stripy phase in the 2D Kitaev-Heisenberg \cite{Chaloupka2013,Rau2014},
thereby providing a quasi-1D understanding to the 2D stripy phase. 

The rest of the paper is organized as follows. 
In Sec. \ref{sec:Ham}, the model Hamiltonian is introduced and the symmetry group is analyzed. 
In Sec. \ref{sec:LL}, the LLRR and $c=1$ phases in the 1D spin-1/2 Kitaev-Heisenberg model are studied by  perturbative Luttinger liquid analysis and DMRG numerics. 
Sec. \ref{sec:KHG_chain} generalizes the analysis to the Kitaev-Heisenberg-Gamma chain,
and shows that the LLRR phase extends to a finite region in the phase diagram of the Kitaev-Heisenberg-Gamma model
when a nonzero Gamma term is introduced. 
Sec. \ref{sec:other_phase} briefly discusses phases in the phase diagram of the 1D Kitaev-Heisenberg-Gamma model other than the LLRR phase,
especially the derivation of the FM phase using a perturbative Luttinger liquid analysis. 
Sec. \ref{sec:stripy} is devoted to a derivation of the stripy phase in the 2D Kitaev-Heisenberg model using a coupled chain method based on the 1D results. 
In Sec. \ref{sec:summary}, we summarize the main results of the paper.

\section{Model Hamiltonian and symmetries}
\label{sec:Ham}

\subsection{Model Hamiltonian}
\label{subsec:Ham}

\begin{figure}[ht]
\begin{center}
\includegraphics[width=8cm]{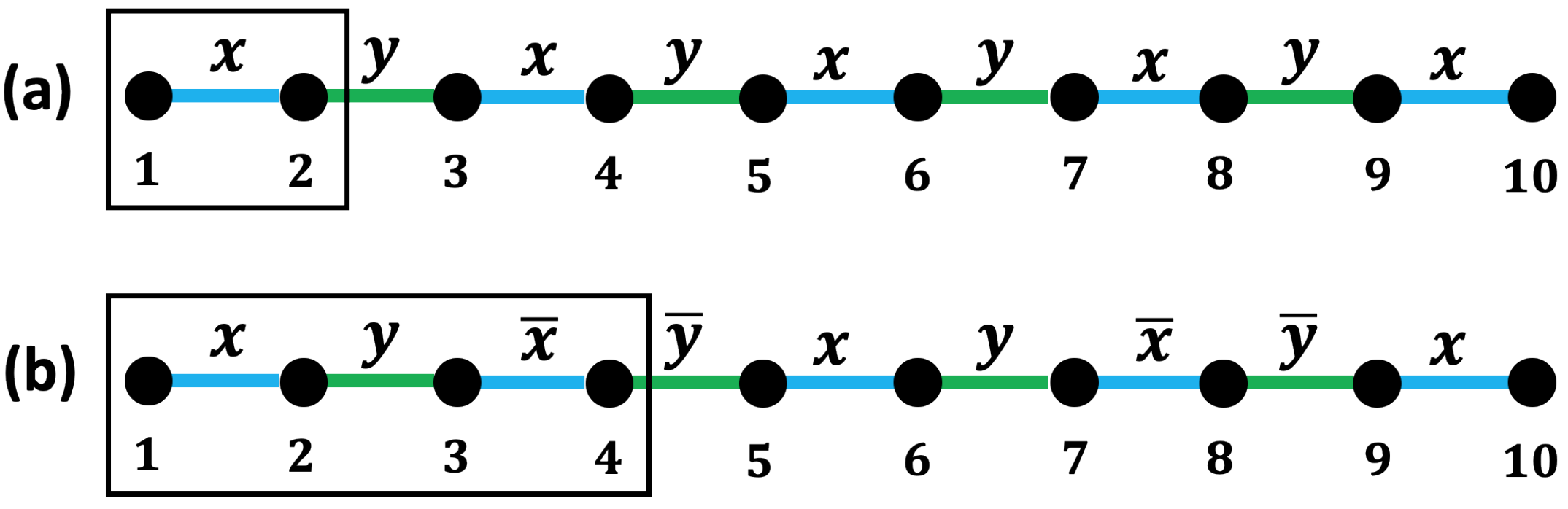}
\caption{Bond pattern of the Kitaev-Heisenberg-Gamma model (a) in the original frame 
and (b) in the four-sublattice rotated frame. 
} \label{fig:bonds}
\end{center}
\end{figure}

We consider the 1D spin-1/2 Kitaev-Heisenberg model defined by the following Hamiltonian,
\begin{flalign}
H_0=\sum_{<ij>\in\gamma\,\text{bond}}\big( KS_i^\gamma S_j^\gamma+ J\vec{S}_i\cdot \vec{S}_j\big),
\label{eq:Ham}
\end{flalign}
in which $i,j$ are two sites of nearest neighbors;
$\gamma=x,y$ is the spin direction associated with the $\gamma$ bond connecting sites $i$ and $j$, with the pattern shown in Fig. \ref{fig:bonds} (a);
$K$ and $J$
are the Kitaev and Heisenberg couplings, respectively.
A useful parametrization is 
 \begin{eqnarray}
 J&=&\cos(\phi),\nn\\ 
 K&=&-\sin(\phi).
\label{eq:parametrization}
\end{eqnarray}
in which $\phi\in[0,2\pi]$.

There is a self-duality transformation $U_4$ of the Kitaev-Heisenberg model called four-sublattice rotation defined by \cite{Chaloupka2013,Chaloupka2015}
\begin{eqnarray}
\text{Sublattice $1$}: & (x,y,z) & \rightarrow (-x^\prime,y^\prime,-z^{\prime}),\nn\\ 
\text{Sublattice $2$}: & (x,y,z) & \rightarrow (-x^{\prime},-y^{\prime},z^{\prime}),\nn\\
\text{Sublattice $3$}: & (x,y,z) & \rightarrow (x^{\prime},-y^{\prime},-z^{\prime}),\nn\\
\text{Sublattice $4$}: & (x,y,z) & \rightarrow (x^{\prime},y^{\prime},z^{\prime}),
\label{eq:4rotation}
\end{eqnarray}
in which ``Sublattice $i$" ($1\leq i \leq 4$) represents all the sites $i+4n$ ($n\in \mathbb{Z}$) in the chain, and the spin operators $S^\alpha$ ($\alpha=x,y,z$) are abbreviated as $\alpha$  for simplicity. 
It can be verified that the Hamiltonian $H^{\prime}_0=U_4 H_0 U_4^{-1}$ in the four-sublattice rotated frame 
is of the same form as Eq. (\ref{eq:Ham}), except that $K$ and $J$ should be replaced by $K+2J$ and $-J$, respectively.
Hence $U_4$ establishes the following equivalence relation in the parameter space,
\bea
(K,J)\simeq (K+2J,-J).
\label{eq:equiv_KJ}
\eea
Particularly, $U_4$ reveals two hidden SU(2) symmetric points $K=-2J<0$  and $K=-2J>0$, corresponding to FM and antiferromagnetic (AFM) Heisenberg models, respectively. 
From here on, we will call the spin coordinate systems  before and after the $U_4$ transformation as the original and $U_4$ frames, respectively. 

Although we mostly focus on the Kitaev-Heisenberg model in this work, 
the spin-1/2 Kitaev-Heisenberg-Gamma model will also be briefly discussed by introducing an additional Gamma term into Eq. (\ref{eq:Ham}).
The Hamiltonian $H_1$ of the Kitaev-Heisenberg-Gamma model is given by
\begin{flalign}
H_1=\sum_{<ij>\in\gamma\,\text{bond}}\big[ KS_i^\gamma S_j^\gamma+ J\vec{S}_i\cdot \vec{S}_j+\Gamma (S_i^\alpha S_j^\beta+S_i^\beta S_j^\alpha)\big],
\label{eq:Ham1_0}
\end{flalign}
in which the pattern of the $\gamma$ bond is shown in Fig. \ref{fig:bonds} (a);
and $\alpha\neq\beta$ are the two remaining spin directions other than $\gamma$.
Since $\Gamma$ changes sign under a global spin rotation around $z$-axis by $\pi$ whereas $K$ and $J$ remain unchanged,
we have the equivalence
\bea
(K,J,\Gamma)\simeq (K,J,-\Gamma).
\label{eq:equiv_Gamma}
\eea
As a result of Eq. (\ref{eq:equiv_Gamma}),  it is enough to consider the $\Gamma>0$ region. 
After the $U_4$ transformation, the Hamiltonian $H^{\prime}_1=U_4 H_1 U_4^{-1}$ becomes
\begin{eqnarray}
H^{\prime}_1&=&\sum_{<ij>\in \gamma\,\text{bond}}\big[ (K+2J)S_i^{\prime\gamma} S_j^{\prime\gamma}-J \vec{S}^\prime_i\cdot \vec{S}^\prime_j \nn\\
&&+\epsilon(\gamma) \Gamma (S_i^{\prime\alpha} S_j^{\prime\beta}+S_i^{\prime\beta} S_j^{\prime\alpha})\big],
\label{eq:4rotated_1}
\end{eqnarray}
in which  the bonds $\gamma=x,y,\bar{x},\bar{y}$ has a four-site periodicity as shown in Fig. \ref{fig:bonds} (b);
the function $\epsilon(\gamma)$ is defined as $\epsilon(x)=\epsilon(y)=-\epsilon(\bar{x})=-\epsilon(\bar{y})=1$;
and $S_i^{\prime\bar{\lambda}}=S_i^{\prime\lambda}$ ($\lambda=x,y,z$).
Notice that in this case, $H^\prime_1$ does not have the same form as $H_1$.

\subsection{Symmetries}
\label{subsec:symmetry_KH}

We give a brief review of the symmetry group structures of $H_0$ and $H_1$, which have been discussed in Ref. \onlinecite{Yang2020}.
In addition, we supplement the analysis in Ref. \onlinecite{Yang2020} with the short exact sequences satisfied by the symmetry groups,
which provide rigorous discussions for the nonsymmorphic structures of the groups. 

\subsubsection{Kitaev-Heisenberg model}

The Kitaev-Heisenberg model in Eq. (\ref{eq:Ham}) in the original frame is invariant under the following symmetry transformations,
\begin{eqnarray}
T &: & (S_i^x,S_i^y,S_i^z)\rightarrow (-S_{i}^x,-S_{i}^y,-S_{i}^z)\nn\\
 T_{2a}&:&  (S_i^x,S_i^y,S_i^z)\rightarrow (S_{i+2}^x,S_{i+2}^y,S_{i+2}^z)\nn\\
T_a I&: &(S_i^x,S_i^y,S_i^z)\rightarrow (S_{-i+1}^x,S_{-i+1}^y,S_{-i+1}^z)\nn\\
R(\hat{y},\pi)&: &(S_i^x,S_i^y,S_i^z)\rightarrow (-S_{i}^x,S_{i}^y,-S_{i}^z)\nn\\
R(\hat{z},-\frac{\pi}{2})T_a&: &(S_i^x,S_i^y,S_i^z)\rightarrow (-S_{i+1}^y,S_{i+1}^x,S_{i+1}^z),
\label{eq:Sym_Neel}
\end{eqnarray}
in which $T$ is the time reversal operation; 
$T_a$ is the translation operator by one site, and
$T_{na}$ is translation by $n$ lattice sites where $n$ is an integer;
$I$ is the spatial inversion with the inversion center located at site $0$;
and $R(\hat{n},\theta)$ denotes the global spin rotation around the $\hat{n}$-direction by an angle $\theta$.
The symmetry group $G_0$ of the Kitaev-Heisenberg chain is
\begin{flalign}
G_0&=\langle
T,T_{2a},T_aI,R(\hat{y},\pi),R(\hat{z},-\frac{\pi}{2})T_a
\rangle,
\label{eq:group_G0}
\end{flalign}
where $\langle...\rangle$ denotes the group generated by the elements within the bracket. 

Since $T_{4a}=[R(\hat{z},-\frac{\pi}{2})T_a]^4$ is a symmetry element and the abelian group $\langle T_{4a}\rangle$ is a normal subgroup of $G_0$, it is legitimate to consider the quotient group $G_0/\langle T_{4a}\rangle$. 
Here we note that one can actually consider $G_0/\langle T_{2a}\rangle$ for the Kitaev-Heisenberg model, but $T_{4a}$ is considered instead so that we can conveniently  compare $G_0/\langle T_{4a}\rangle$ with the symmetry group of the Kitaev-Heisenberg-Gamma model in the $U_4$ frame to be discussed later. 

\begin{figure}[ht]
\begin{center}
\includegraphics[width=5cm]{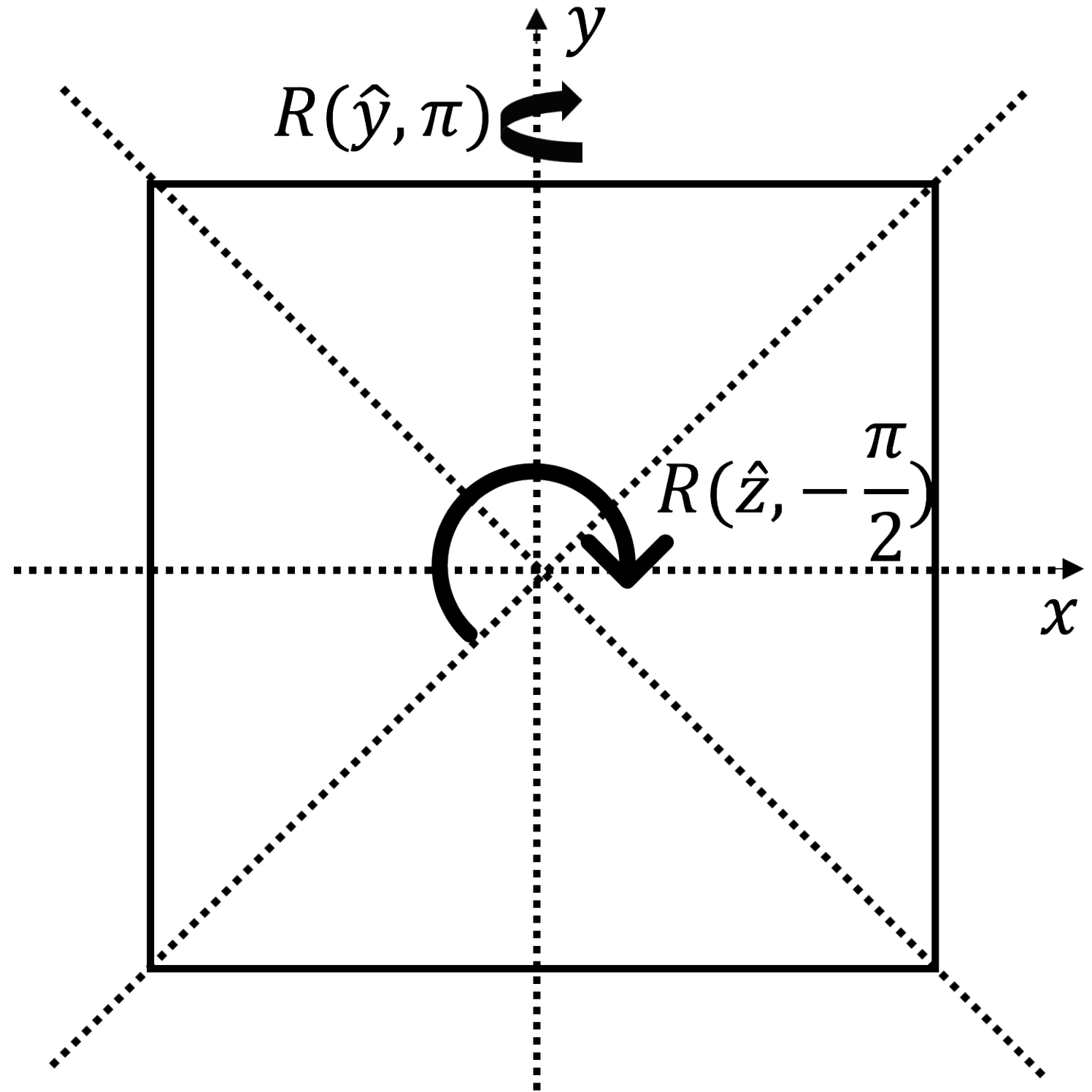}
\caption{$R(\hat{y},\pi)$ and $R(\hat{z},-\pi/2)$ as symmetry operations of a square in the spin space,
in which the $z$-direction is perpendicular to the plane.
The figure is taken from Ref. \onlinecite{Yang2020}.
} 
\label{fig:D4d}
\end{center}
\end{figure}

It has been proved in Ref. \onlinecite{Yang2020} that $G_0/\langle T_{4a}\rangle$ is isomorphic to $(\mathbb{Z}_2\times \mathbb{Z}_2) \ltimes D_{4h}$,
in which 
$\mathbb{Z}_2\times\mathbb{Z}_2=\langle T_aI,T_{2a}\rangle/\langle T_{4a}\rangle$
and $D_{4h}=\langle T\rangle\times D_4$, where $D_4$ is the dihedral group of order $8$ given by $D_4=\langle R(\hat{z},-\frac{\pi}{2})T_a,R(\hat{y},\pi)T_a I
\rangle /\langle T_{4a}\rangle$.
We note that an intuitive  way to understand the origin of the $D_4$ group is by observing that if the spatial operations $T_a$ and $T_aI$ are removed  from $R(\hat{z},-\frac{\pi}{2})T_a$ and $R(\hat{y},\pi)T_a I$,
then the group $\langle R(\hat{z},-\frac{\pi}{2}),R(\hat{y},\pi)\rangle$  generated by actions in the spin space represents  the symmetry group of a spin square shown in Fig. \ref{fig:D4d}, which is exactly the $D_4$ group.
As a consequence of the above analysis of the group structure of $G_0$, we have the short exact sequence
\bea
1\rightarrow \langle T_{4a}\rangle\rightarrow G_0\rightarrow (\mathbb{Z}_2\times \mathbb{Z}_2) \ltimes D_{4h}\rightarrow 1.
\label{eq:short_exact_0}
\eea
Particularly, the group $G_0$ is nonsymmorphic in the sense that the short exact sequence in Eq. (\ref{eq:short_exact_0}) is non-splitting,
i.e., there is no way to embed the quotient group $G_0/\langle T_{4a}\rangle$ into $G_0$.

\subsubsection{Kitaev-Heisenberg-Gamma model}
\label{subsec:symmetry_KHG}

In Ref. \onlinecite{Yang2020}, it is further shown that the symmetry group $G_1$ of the Kitaev-Heisenberg-Gamma model in Eq. (\ref{eq:Ham1_0}) in the original frame is 
\begin{eqnarray}
G_1=\langle T,T_aI,R(\hat{n}_1,\pi)T_a
\rangle,
\label{eq:GN}
\end{eqnarray}
in which $\hat{n}_1=\frac{1}{\sqrt{2}}(1,-1,0)^T$.
The group $G_1$ satisfies $G_1/\langle T_{4a}\rangle=\mathbb{Z}_2\ltimes C_{4h}$,
in which  $\mathbb{Z}_2=\langle T_a I\rangle$ and $C_{4h}=\langle T\rangle \times C_4$ where $C_4=\langle R(\hat{n}_1,\pi)T_a\rangle/\langle T_{4a}\rangle$ is the cyclic group of order $4$.
Therefore, there is the short exact sequence 
\bea
1\rightarrow \langle T_{4a}\rangle\rightarrow G_1\rightarrow \mathbb{Z}_2 \ltimes C_{4h}\rightarrow 1,
\label{eq:short_exact_1}
\eea
and $G_1$ is again a nonsymmorphic group since the above short exact sequence is non-splitting.
Clearly, the quotient $\mathbb{Z}_2 \ltimes C_{4h}$  in Eq. (\ref{eq:short_exact_1}) is a subgroup of $(\mathbb{Z}_2\times \mathbb{Z}_2) \ltimes D_{4h}$ in Eq. (\ref{eq:short_exact_0}), 
as $G_1$ is a subgroup of $G_0$.

On the other hand, the symmetry group  $G^\prime_1$ of the Kitaev-Heisenberg-Gamma model in the $U_4$ frame in Eq. (\ref{eq:4rotated_1}) is given by \cite{}
\begin{flalign}
G^\prime_1&=\langle
T,R(\hat{y},\pi)T_aI,R(\hat{z},-\frac{\pi}{2})T_a
\rangle,
\label{eq:group_G1}
\end{flalign}
which  is a nonsymmorphic group and satisfies
\bea
1\rightarrow \langle T_{4a}\rangle\rightarrow G^\prime_1\rightarrow   D_{4h}\rightarrow 1.
\label{eq:short_exact_1_prime}
\eea
It is straightforward to see that $G_1^\prime$ is a subgroup of $G_0$ by comparing Eq. (\ref{eq:short_exact_1_prime}) with Eq. (\ref{eq:short_exact_0}),
where $D_{4h}$ is a subgroup of $(\mathbb{Z}_2\times \mathbb{Z}_2) \ltimes D_{4h}$.
We note that $G_1$ and $G_1^\prime$ are isomorphic to each other
since they are the symmetry groups of the same model albeit in different frames.
This is indeed true, as can be seen from the isomorphism  $\mathbb{Z}_2 \ltimes C_{4h}\cong D_{4h}$. 

\subsubsection{Nonsymmorphic nature of the symmetry groups}

For all above nonsymmorphic symmetry groups in Sec. \ref{subsec:symmetry_KH} and Sec. \ref{subsec:symmetry_KHG}, 
the non-splitting property can be most easily seen by considering the subgroup $\langle R(\hat{z},-\frac{\pi}{2})T_a\rangle$ and the following short exact sequence,
\bea
1\rightarrow \langle T_{4a}\rangle\rightarrow \langle R(\hat{z},-\frac{\pi}{2})T_a\rangle\rightarrow C_4\rightarrow 1,
\label{eq:short_exact_sub}
\eea
in which $C_4\cong \langle R(\hat{z},-\frac{\pi}{2})T_a\rangle /\mathopen{<}T_{4a}\mathclose{>}$.
Replacing the groups in Eq. (\ref{eq:short_exact_sub}) with their isomorphic counterparts, Eq. (\ref{eq:short_exact_sub}) can be alternatively written as
\bea
1\rightarrow 4\mathbb{Z}\rightarrow \mathbb{Z}\rightarrow \mathbb{Z}_4\rightarrow 1,
\eea
which is clearly non-splitting (namely, $\mathbb{Z}_4$ cannot be viewed as a subgroup of $\mathbb{Z}$ in the sense of embedding), where $\mathbb{Z}_4=\mathbb{Z}/4\mathbb{Z}$.

\subsection{Phase diagram}
\label{subsec:phase}

Before going to technical discussions, we give a description  of the phase diagram shown in Fig. \ref{fig:phase}.

\subsubsection{Kitaev-Heisenberg model}

We first describe  the phase diagram of the Kitaev-Heisenberg model corresponding to the horizontal axis in  Fig. \ref{fig:phase}.
The AFM Heisenberg point $J$, FM Heisenberg point $-J$, and the FM Kitaev point $-K$ are located at $\phi=0$, $\phi=\pi$, and $\phi=\pi/2$, respectively.
There is another special point FM$^\prime$ at $\phi=\arctan(2)$, which is a hidden SU(2) symmetric point satisfying $K=-2J<0$.
We note that  the region $\phi\in(\arctan(2),\pi/2)$ is related to $\phi\in(\pi/2,\pi)$ by the $U_4$ transformation, hence the physics in these two regions are unitarily  equivalent. 

The horizontal axis in  Fig. \ref{fig:phase} divides into five phases:
AFM phase between $J$ and FM$^\prime$;
LLRR phase between FM$^\prime$ and $\phi_{c1}\approx 0.4\pi$;
a gapless $c=1$ phase between $\phi_{c1}$ and $-K$, where $c$ represents the central charge;
another  $c=1$ phase between $-K$ and $\phi_{c2}\approx 0.56 \pi$, which is dual to the $c=1$ phase to the left of the $-K$ point in Fig. \ref{fig:phase} under the $U_4$ transformation; 
and an FM phase between $\phi_{c2}$ and $-J$, which is dual to the LLRR phase in the region $\phi\in[\text{FM}^\prime,\phi_{c1}]$ under the $U_4$ transformation. 

The spin ordering has a four-site periodicity in the LLRR phase in the original frame, 
aligning along $\hat{x}$, $\hat{x}$, $-\hat{x}$, $-\hat{x}$
or $\hat{y}$, $\hat{y}$, $-\hat{y}$, $-\hat{y}$ directions in a four-site unit cell,
hence the name LLRR (left-left-right-right for short). 
We note that the LLRR phase was erroneously identified as a spiral phase in Ref. \onlinecite{Yang2020}. 
The numerics become very difficult in the $c=1$ phase, 
since the system is very close to the exactly solvable FM Kitaev point, which  has an exponentially large infinite ground state degeneracy in the thermodynamic limit (degeneracy is $2^{L/2}$ for a system of $L$ sites) \cite{Brzezicki2007,You2008}.
More detailed analytical and numerical analysis of the $c=1$ phase will be left for future investigations. 

\subsubsection{Kitaev-Heisenberg-Gamma model}

By adding a small Gamma interaction, nearly all the phases extend to a finite area in the two-parameter phase diagram as shown in Fig. \ref{fig:phase}.
In particular, the LLRR phase on the $\Gamma=0$ line extends to the d-LLRR phase, 
where ``d-LLRR" is ``distorted-left-left-right-right" for short,
the name of which comes from the fact that the pattern of the spin ordering can be viewed as a slightly distorted version of the LLRR ordering in Kitaev-Heisenberg model, no longer collinear when the Gamma term is nonzero. 
On the other hand, whether the gapless $c=1$ phase on the horizontal axis extends to a finite region remains unclear,
which is denoted as the ``KSL" (``Kitaev spin liquid" for short) phase in Fig. \ref{fig:phase} and will be left for future studies.
When $\Gamma$ becomes larger, there appears a Luttinger liquid phase in the positive $J$ region denoted as ``LL" in Fig. \ref{fig:phase},
which has been discussed in detail in Ref. \onlinecite{Yang2020}.

\section{Perturbative Luttinger liquid analysis of the Kitaev-Heisenberg model}
\label{sec:LL}

In this section, we provide an analytical understanding for the origin of the LLRR phase in the spin-1/2 Kitaev-Heisenberg chain based on  a perturbative Luttinger liquid analysis. 
We will work in the $U_4$ frame unless otherwise stated. 
Since the region between $-K$ and $-J$ is dual to the region between FM$^\prime$ and $-K$ via the $U_4$ transformation,
the perturbative Luttinger liquid analysis can also be applied to explain the FM phase for $\phi\in[\phi_{c2},\pi]$.

\subsection{Failure of the classical analysis}
\label{subsec:classical}

The simplest approach is the classical analysis by treating the spin operators as vectors of real numbers which is valid  in the large-$S$ limit. 
We will show that the classical analysis leads to an infinitely degenerate ground state manifold,
and fails since it ignores the quantum fluctuations.  

In the classical analysis, the spin operator $\vec{S}_i$ is approximated as
\bea
\vec{S}_i=S(x_i,y_i,z_i)^T,
\eea
in which $S$ is the value of the spin, and $x_i,y_i,z_i$ are normalized as
\bea
(x_i)^2+(y_i)^2+(z_i)^2=1.
\label{eq:constraint_unit}
\eea
Assuming that the ground state has a two-site periodicity, i.e., $\vec{S}^\prime_{i+2}=\vec{S}^\prime_i$,
the classical free energy per two-site unit cell in the $U_4$ frame becomes
\bea
f&=&(K+2J)S^2(x^\prime_1x^\prime_2+y^\prime_1y^\prime_2)-2JS^2\hat{n}^\prime_1\cdot\hat{n}^\prime_2\nn\\
&&-\frac{1}{2}\lambda_1 (\hat{n}_1^\prime\cdot\hat{n}_1^\prime-1)-\frac{1}{2}\lambda_2 (\hat{n}^\prime_2\cdot\hat{n}^\prime_2-1),
\label{eq:classical_f}
\eea
in which $\hat{n}^\prime_i=(x^\prime_i,y^\prime_i,z^\prime_i)$, and $\lambda_i$ ($i=1,2$) are introduced as a Lagrange multiplier to impose the constraints in Eq. (\ref{eq:constraint_unit}). 
Eq. (\ref{eq:classical_f}) can be rewritten as
\bea
f&=&(-2J+\delta) S^2(x^\prime_1x^\prime_2+y^\prime_1y^\prime_2)-2J S^2 z^\prime_1z^\prime_2-\nn\\
&&\frac{1}{2}\lambda_1 (\hat{n}^\prime_1\cdot\hat{n}^\prime_1-1)-\frac{1}{2}\lambda_2 (\hat{n}^\prime_1\cdot\hat{n}^\prime_2-1),
\label{eq:classical_f2}
\eea
in which $\delta=K+2J$.
In the region between FM$^\prime$ and $-K$ on the horizontal axis in Fig. \ref{fig:phase}, we have $J>0$ and $\delta<0$.
Therefore, an FM configuration within the $xy$-plane is able to lower the energy in Eq. (\ref{eq:classical_f2}).
However, it is clear that the free energy $f$ in Eq. (\ref{eq:classical_f2}) has a U(1) symmetry, corresponding to a rotational symmetry around the $z$-axis.
Hence, the classical ground states are infinitely degenerate, since the FM spin direction can point to any direction within the $xy$-plane. 
This is clearly absurd, since the system only has discrete symmetries and cannot develop a symmetry breaking pattern which has a continuously variable spin orientation.  

Quantum fluctuations will break the infinite classical ground state degeneracy.
There are two possibilities though: 
The system either becomes a gapless Luttinger liquid (like the easy-plane XXZ model where the classical analysis also gives an infinite ground state degeneracy but the system is gapless at the quantum level),
or is magnetically ordered but having a discrete spontaneous  symmetry breaking pattern. 
We will show that at quantum level, the Luttinger liquid behavior is unstable under the symmetry allowed perturbations, 
and as a result, the system develops a magnetic order with a discrete symmetry breaking pattern. 

\subsection{Perturbative Luttinger liquid analysis}
\label{subsec:Luttinger}

In the $U_4$ frame, we write the Hamiltonian in the region $\phi\in(\arctan{2},\pi/2)$ as
\bea
H_0^\prime= H^\prime_{XXZ}+\frac{1}{2}\delta \sum_i (-)^{i-1} (S_i^{\prime x}S_{i+1}^{\prime x}-S_i^{\prime y}S_{i+1}^{\prime y}), 
\label{eq:H0_delta}
\eea
in which 
\begin{flalign}
H^\prime_{XXZ}=\sum_i[ (-J+\frac{1}{2}\delta)(S_i^{\prime x}S_{i+1}^{\prime x}+S_i^{\prime y}S_{i+1}^{\prime y})-JS_i^{\prime z}S_{i+1}^{\prime z}],
\label{eq:Hp_XXZ}
\end{flalign}
and 
\bea
\delta=K+2J<0. 
\eea 
It can be seen that $H^\prime_{XXZ}$ represents an easy-plane XXZ model when $\delta<0$,
hence lies in the gapless Luttinger liquid phase \cite{}.
The strategy is to take the second term in Eq. (\ref{eq:H0_delta}) as a perturbation and analyze its effects using the Luttinger liquid theory. 
We emphasize that the problem is non-perturbative in nature, since the coupling constant $\frac{1}{2}\delta$  in the second term of $H_0^\prime$ is comparable with the difference between the transverse and longitudinal couplings in $H^\prime_{XXZ}$ in Eq. (\ref{eq:Hp_XXZ}).
Therefore, the perturbative Luttinger liquid analysis to be discussed below is not analytically  controllable in the most rigorous sense, and must be verified by numerical simulations.

For later convenience, we further consider a two-sublattice rotation $V_2$ defined as
\bea
\text{Sublattice $1$}: & (x^\prime,y^\prime,z^\prime) & \rightarrow (x^{\prime\prime},y^{\prime\prime},z^{\prime\prime}),\nn\\ 
\text{Sublattice $2$}: & (x^\prime,y^\prime,z^\prime) & \rightarrow (-x^{\prime\prime},-y^{\prime\prime},z^{\prime\prime}),
\label{eq:V_2}
\eea
in which ``Sublattice $i$" ($1\leq i \leq 2$) represents all the sites $i+2n$ ($n\in \mathbb{Z}$) in the chain, and the spin symbol $S$ in $S^\alpha$ is dropped for simplicity.
Then we obtain 
\bea
H_0^{\prime\prime}= H^{\prime\prime}_{XXZ}+H_\delta, 
\label{eq:H0_delta_2}
\eea
in which
\bea
H^{\prime\prime}_{XXZ}&=&(J-\frac{1}{2}\delta) \sum_i[ S_i^{\prime\prime x}S_{i+1}^{\prime\prime x}+S_i^{\prime\prime y}S_{i+1}^{\prime\prime y}+\Delta S_i^{\prime z}S_{i+1}^{\prime\prime z}],\nn\\
H_\delta&=&\frac{1}{2}\delta \sum_i (-)^{i} (S_i^{\prime\prime x}S_{i+1}^{\prime\prime x}-S_i^{\prime\prime y}S_{i+1}^{\prime\prime y}),
\label{eq:H_Delta_delta}
\eea
where 
\bea
\Delta=-\frac{J}{J-\frac{1}{2}\delta}.
\eea
The advantage of applying $V_2$ is that the correlation functions of $H^{\prime\prime}_{XXZ}$ in the $xy$-plane are rendered to be dominated by AFM rather than FM fluctuations,
which is the commonly used convention in Luttinger liquid theory. 
After the application of  the $V_2$ transformation, the symmetry operations in Eq. (\ref{eq:Sym_Neel}) become
\begin{eqnarray}
V_2T(V_2)^{-1} &= &T \nn\\
V_2T_{2a}(V_2)^{-1}&=& T_{2a} \nn\\
V_2T_a I(V_2)^{-1}&= & R(\hat{z}^{\prime\prime},\pi) T_a I \nn\\
V_2R(\hat{y}^{\prime\prime},\pi)(V_2)^{-1}&= &R(\hat{y}^{\prime\prime},\pi)\nn\\
V_2R(\hat{z}^{\prime\prime},-\frac{\pi}{2})T_a(V_2)^{-1}&=&R(\hat{z}^{\prime\prime},\frac{\pi}{2})T_a.
\label{eq:Sym_Neel_V2}
\end{eqnarray}

The easy-plane XXZ model $H^{\prime\prime}_{XXZ}$ for $\delta<0$ can be bosonized in the standard way \cite{}.
The low energy physics of $H^{\prime\prime}_{XXZ}$ is described by the following Luttinger liquid theory,
\bea
H_{LL}=\frac{v}{2} \int dx [\frac{1}{\kappa} (\nabla \varphi)^2 +\kappa (\nabla \theta)^2],
\label{eq:LL_liquid}
\eea
in which $v$ is the velocity, $\kappa$ is the Luttinger liquid parameter,
and the $\theta$ and $\varphi$ fields satisfy the commutation relation $[\varphi(x),\theta(x^\prime)]=\frac{i}{2}\text{sgn}(x^\prime-x)$.
The local spin operators of the XXZ model is related to the boson fields $\theta,\varphi$ via the following bosonization formulas
 \begin{flalign}
S^{\prime\prime z}(x)&= -\frac{1}{\sqrt{\pi}} \nabla \varphi(x) + \text{const.} \frac{1}{a}(-)^n \cos(2\sqrt{\pi} \varphi(x)),\nn\\
S^{\prime\prime +}(x)&=\text{const.} \frac{1}{\sqrt{a}}e^{-i\sqrt{\pi}\theta(x)} \big[(-)^n+\cos(2\sqrt{\pi}\varphi(x))\big],
\label{eq:abelian_bosonize}
\end{flalign}
in which $S^{\prime\prime +}=S^{\prime\prime x}+iS^{\prime\prime y}$,
and $x=na$ ($n\in \mathbb{Z}$) is the spatial  coordinate in the continuum limit.

The easiest way to treat the perturbation $H_\delta$ is to do a first order perturbation, namely,
rewriting $H_{LL}$ in terms of the $\theta,\varphi$ fields by replacing the spin operators in $H_\delta$ with the expressions in Eq. (\ref{eq:abelian_bosonize}).
The first order projection can be evaluated as
\begin{flalign}
&\sum_i (-)^i(S_i^{\prime\prime x}S_{i+1}^{\prime\prime x}-S_i^{\prime\prime y}S_{i+1}^{\prime\prime y})=\nn\\
&\frac{(\text{const.})^2}{a^2}\int dx \big[\cos(\sqrt{\pi} \theta(x))\cos(\sqrt{\pi} \theta(x+a))\nn\\
&-\sin(\sqrt{\pi} \theta(x))\sin(\sqrt{\pi} \theta(x+a))\big]\nn\\
&\times\big[\cos(2\sqrt{\pi}\varphi(x+a))-\cos(2\sqrt{\pi}\varphi(x))\big]+\cdot \cdot \cdot,
\label{eq:first_order_projection}
\end{flalign}
in which ``$\cdot\cdot\cdot$" denotes the terms not included in the first order projection. 
Eq. (\ref{eq:first_order_projection}) can be simplified using the operator product expansions of vertex operators.
However, we don't need to bother ourselves to perform such simplifications, since 
the operators in Eq. (\ref{eq:first_order_projection}) are highly irrelevant in the vicinity of the FM$^\prime$ point in the sense of renormalization group (RG).
To see this, notice that the scaling dimension of $\cos(2\sqrt{\pi} \varphi)$ is $\kappa$,
where $\kappa$ is the Luttinger parameter. 
However, $\kappa$ diverges when the easy-plane XXZ model approaches the FM limit,
i.e., when $\Delta\rightarrow-1$ ($|\Delta|<1$) in Eq. (\ref{eq:H_Delta_delta}).
This is exactly the case when $\delta$ in Eq. (\ref{eq:H_Delta_delta}) is small.
Hence we see that when $|\delta|\ll 1$, Eq. (\ref{eq:first_order_projection}) has a very large scaling dimension,
therefore can be neglected in the Luttinger liquid theory when low energy physics is considered. 

The above analysis shows that we need to go beyond the first order projection.
Since such calculations are difficult, we will not explicitly derive the high order perturbations.
Instead, we use a symmetry analysis to figure out the   most relevant symmetry allowed terms in the RG sense.
It is worth to note that although the symmetry analysis is able to determine the form of the corresponding relevant operator, 
the signs and order of magnitudes of the coupling constants cannot be figured out.
For our purpose, the order of magnitude of the coupling constant is not important,
whereas the sign is crucial which determines the specific type of magnetic instability. 
We will infer the sign of the coupling constant indirectly  by numerical calculations in later sections. 

To perform the symmetry analysis, the transformation properties of the $\theta,\varphi$ fields are needed.
Such transformations can be derived by first performing a Jordan-Wigner transformation to the spin operators, then bosonizing the obtained  spinless fermion.
The results are
 (for details, see Appendix \ref{app:symmetry_transform})
\begin{flalign}
&T:  \theta(t,x)\rightarrow \theta(-t,x)+\sqrt{\pi}, \varphi(t,x)\rightarrow -\varphi(-t,x),\nn\\
&~~~~~\eta_r^\dagger\rightarrow \eta_{-r},\nn\\
&T_a: \theta\rightarrow \theta +\sqrt{\pi},\varphi\rightarrow \varphi+\sqrt{\pi}/2,\eta_r^\dagger\rightarrow \eta^\dagger_{r},\nn\\
&I : \theta(t,x)\rightarrow \theta(t,-x),\varphi(t,x)\rightarrow -\varphi(t,-x)+\sqrt{\pi}S_T^z,\nn\\
&~~~~~\eta_r^\dagger\rightarrow \eta^\dagger_{-r},\nn\\
&R(\hat{y}^{\prime\prime},\pi):\theta\rightarrow -\theta +\sqrt{\pi},\varphi\rightarrow -\varphi,\eta_r^\dagger\rightarrow \eta_r,\nn\\
&R(\hat{z}^{\prime\prime},\beta): \theta\rightarrow \theta+\beta/\sqrt{\pi},\varphi\rightarrow \varphi,\eta_r^\dagger\rightarrow \eta^\dagger_{r},
\label{eq:transformation_theta_phi}
\end{flalign}
where $\eta_r$ ($r=L,R$) are the Klein factors for the left and right movers of the spinless fermion,
and $S_T^z=\sum_{k\in \mathbb{Z}} S_k^z$ is the total spin along $z$-direction, which is a good quantum number for the XXZ chain. 

Next we analyze what vertex operators are allowed by symmetries.
First notice that only operators of the form $e^{i\sqrt{\pi}(2n \varphi+m\theta)}$ are allowed,
since the bosonization formulas for the spin operators in Eq. (\ref{eq:abelian_bosonize})
only contain $\pm2\sqrt{\pi}\varphi,\pm\sqrt{\pi}\theta$ in the exponentials,
and a perturbation, whatever its order, can only give integer multiples of $2\sqrt{\pi}\varphi$ and $\sqrt{\pi}\theta$. 
Since the scaling dimension of $e^{2i\lambda \sqrt{\pi}\varphi}$ ($\lambda\in\mathbb{Z}$) diverges when $|\delta|\ll 1$, 
we consider $e^{im\sqrt{\pi}\theta}$,
which are the only possible relevant vertex operators at low energies.

Using Eq. (\ref{eq:transformation_theta_phi}), 
the transformations of $\cos(m\sqrt{\pi}\theta)$ and $\sin(m\sqrt{\pi}\theta)$ ($m\in \mathbb{Z}$)  under the symmetries operations in Eq. (\ref{eq:Sym_Neel}) can be derived as
\bea
T&:& \cos(m\sqrt{\pi}\theta)\rightarrow (-)^m \cos(m\sqrt{\pi}\theta),\nn\\
&&\sin(m\sqrt{\pi}\theta)\rightarrow (-)^m \sin(m\sqrt{\pi}\theta),
\label{eq:theta_T}
\eea
\bea
T_{2a}&:& \cos(m\sqrt{\pi}\theta)\rightarrow \cos(m\sqrt{\pi}\theta),\nn\\
&&\sin(m\sqrt{\pi}\theta)\rightarrow \sin(m\sqrt{\pi}\theta),
\label{eq:theta_T2a}
\eea
\bea
R(\hat{z}^{\prime\prime},\pi)T_aI&:&\cos(m\sqrt{\pi}\theta(x))\rightarrow \cos(m\sqrt{\pi}\theta(-x)),\nn\\
&&\sin(m\sqrt{\pi}\theta(x))\rightarrow \sin(m\sqrt{\pi}\theta(-x)),
\label{eq:theta_TaI}
\eea
\bea
R(\hat{y}^{\prime\prime},\pi)&:& \cos(m\sqrt{\pi}\theta)\rightarrow (-)^m\cos(m\sqrt{\pi}\theta),\nn\\
&&\sin(m\sqrt{\pi}\theta)\rightarrow (-)^{m+1}\sin(m\sqrt{\pi}\theta),
\label{eq:theta_Ry}
\eea
\bea
R(\hat{z}^{\prime\prime},\frac{\pi}{2})T_a&: &\cos(m\sqrt{\pi}\theta)\rightarrow (-)^m\cos(m\sqrt{\pi}\theta+\frac{m\pi}{2}),\nn\\
&&\sin(m\sqrt{\pi}\theta)\rightarrow (-)^m\sin(m\sqrt{\pi}\theta+\frac{m\pi}{2}).\nn\\
\label{eq:theta_Rz}
\eea
Combining Eq. (\ref{eq:theta_T}) with Eq. (\ref{eq:theta_Ry}),
it can be seen that $\sin(m\sqrt{\pi}\theta)$ is forbidden,
and $m$ has to be an even integer for $\cos(m\sqrt{\pi}\theta)$.
Then Eq. (\ref{eq:theta_Rz}) requires $\cos(4n\sqrt{\pi}\theta)$,
which is invariant under all symmetry transformations.
Taking the smallest possible number $n=1$ (which has the smallest scaling dimension),
the most relevant symmetry allowed perturbation is given by
\bea
g\int dx \cos(4\sqrt{\pi}\theta),
\label{eq:cos4theta}
\eea
in which $g$ is the coupling constant. 
The operator in Eq. (\ref{eq:cos4theta}) has a scaling dimension $4/\kappa$,
which is highly relevant when $\kappa$ becomes large. 
In fact, $\kappa$ can be arbitrarily large in the vicinity of the FM$^\prime$ point on the horizontal axis in Fig. (\ref{fig:phase}).

\subsection{Magnetic ordering}
\label{subsec:order}

Next, we figure out the magnetic order arising from the operator in Eq. (\ref{eq:cos4theta}).
We need to distinguish between two cases, i.e., $g>0$ and $g<0$.

We first consider $g>0$. 
In this case, the energy in Eq. (\ref{eq:cos4theta}) is minimized at $4\sqrt{\pi}\theta=(2n+1)\pi$, i.e.,
\bea
\theta^{(\text{I})}_n=\frac{2n+1}{4}\sqrt{\pi},~0\leq n\leq 3,
\label{eq:sol_theta_I}
\eea
where the superscript ``$(\text{I})$" is used  to denote the case for  $g>0$.
We take one of the four degenerate solutions at $n=0$ as an example,
giving  $\theta_0^{(\text{I})}=\frac{\sqrt{\pi}}{4}$.
Plugging $\theta_0^{(\text{I})}$ into the bosonization formulas in Eq. (\ref{eq:abelian_bosonize}), the spin orientations can be determined as
\bea
\vec{S}^{(\text{I})\prime\prime}_j&=&(-)^j\frac{1}{\sqrt{2}}(1,1,0)^T.
\label{eq:order_S_pprime}
\eea
Performing $(V_2)^{-1}$ to Eq. (\ref{eq:order_S_pprime}),
we obtain
\bea
\vec{S}^{(\text{I})\prime}_j&=&\frac{1}{\sqrt{2}}(1,1,0)^T.
\label{eq:order_S_prime}
\eea

It is useful to determine the symmetry breaking pattern of Eq. (\ref{eq:order_S_prime}).
It can be verified that Eq. (\ref{eq:order_S_prime}) is invariant under the following symmetry operations 
\begin{flalign}
&H^{(\text{I})}=\nn\\
&\langle T_aI,T_{2a},[R(\hat{z},-\frac{\pi}{2})T_a]^2T,R(\hat{z},-\frac{\pi}{2})T_a\cdot R(\hat{y},\pi)T_aI\rangle,
\label{eq:HI}
\end{flalign}
in which $H^{(\text{I})}$ is the unbroken symmetry group. 
Using a similar analysis as Ref. \onlinecite{Yang2020},  
$H^{(\text{I})}$ can be shown to satisfy 
\bea
1\rightarrow \langle T_{4a}\rangle \rightarrow H^{(\text{I})}\rightarrow (\mathbb{Z}_2\times \mathbb{Z}_2)\ltimes D^{(\text{I})}_2 \rightarrow 1,
\label{eq:short_exact_KH_spiral}
\eea
in which 
\bea
\mathbb{Z}_2\times \mathbb{Z}_2&=&\langle T_aI,T_{2a}\rangle/\langle T_{4a}\rangle,
\eea
and
\begin{flalign}
&D^{(\text{I})}_2=\nn\\
&\langle [R(\hat{z},-\frac{\pi}{2})T_a]^2T,R(\hat{z},-\frac{\pi}{2})T_a\cdot R(\hat{y},\pi)T_aI\rangle /\langle T_{4a}\rangle,
\label{eq:D_2_I}
\end{flalign}
where $D_2^{(\text{I})}\cong D_2$ is the dihedral group of order $4$.
Combining Eq. (\ref{eq:short_exact_0}) with Eq. (\ref{eq:HI}), we see that the symmetry breaking pattern is
\bea
(\mathbb{Z}_2\times \mathbb{Z}_2)\ltimes D_{4h}\rightarrow (\mathbb{Z}_2\times \mathbb{Z}_2)\ltimes D_2^{\text{(I)}}.
\eea
Since $|[(\mathbb{Z}_2\times \mathbb{Z}_2)\ltimes D_{4h}]/[(\mathbb{Z}_2\times \mathbb{Z}_2)\ltimes D_2]|=4$,
there are four degenerate configurations, consistent with the analysis in Eq. (\ref{eq:sol_theta_I}).

The most general magnetic pattern can be figured out by requiring invariance under $H^{(\text{I})}$, which gives
\bea
\vec{S}^{(\text{I})\prime}_j&=&N_1\frac{1}{\sqrt{2}}(1,1,0)^T,
\label{eq:order_S_prime_general}
\eea
in which $N_1$ is the magnitude of the spin ordering (for derivations, see Appendix \ref{app:determine_orde1}).
Rotating back to the original frame by performing $(U_4)^{-1}$, we obtain
\bea
\vec{S}^{(\text{I})}_{1+4n}&=&N_1\frac{1}{\sqrt{2}}(-1,1,0)^T,\nn\\
\vec{S}^{(\text{I})}_{2+4n}&=&N_1\frac{1}{\sqrt{2}}(-1,-1,0)^T,\nn\\
\vec{S}^{(\text{I})}_{3+4n}&=&N_1\frac{1}{\sqrt{2}}(1,-1,0)^T,\nn\\
\vec{S}^{(\text{I})}_{4+4n}&=&N_1\frac{1}{\sqrt{2}}(1,1,0)^T.
\label{eq:original_0}
\eea
This is a spiral phase in which the spins rotate in the $xy$-plane as the position is moved along the chain. 
The pattern of the magnetic ordering for the other three degenerate ground states can be obtained by applying the broken symmetry transformations to Eq. (\ref{eq:original_0}). 

Next we consider the $g<0$ case. 
The energy in Eq. (\ref{eq:cos4theta}) is minimized at $4\sqrt{\pi}\theta=2n\pi$, i.e.,
\bea
\theta^{(\text{II})}_n=\frac{n}{2}\sqrt{\pi},~0\leq n\leq 3. 
\label{eq:theta_II}
\eea
We take one of the four degenerate solutions at $n=1$ as an example,
giving $\theta_0^{(\text{II})}=\frac{\pi}{2}$.
Plugging $\theta_0^{(\text{II})}$ into the bosonization formulas in Eq. (\ref{eq:abelian_bosonize}), the spin orientations can be determined as
\bea 
\vec{S}^{(\text{II})\prime\prime}_j&=&(-)^j(0,1,0)^T.
\label{eq:order_S_pprime_II}
\eea
Performing $(V_2)^{-1}$ to Eq. (\ref{eq:order_S_pprime_II}),
we obtain
\bea
\vec{S}^{(\text{II})\prime}_j&=&\frac{1}{\sqrt{2}}(0,1,0)^T.
\label{eq:order_S_prime_II}
\eea
It can be verified that Eq. (\ref{eq:order_S_prime_II}) is invariant under the following symmetry operations 
\bea
H^{(\text{II})}=\langle T_aI,T_{2a},[R(\hat{z},-\frac{\pi}{2})T_a]^2T,R(\hat{y},\pi)T_aI\rangle,
\label{eq:HII}
\eea
in which $H^{(\text{II})}$ is the unbroken symmetry group. 
Similarly, $H^{(\text{II})}$ satisfies
\bea
1\rightarrow \langle T_{4a}\rangle \rightarrow H^{(\text{II})}\rightarrow (\mathbb{Z}_2\times \mathbb{Z}_2)\ltimes D^{(\text{II})}_2 \rightarrow 1,
\eea
in which 
\bea
\mathbb{Z}_2\times \mathbb{Z}_2=\langle T_aI,T_{2a}\rangle/\langle T_{4a}\rangle,
\eea
and
\bea
D^{(\text{II})}_2=\langle [R(\hat{z},-\frac{\pi}{2})T_a]^2T,R(\hat{y},\pi)T_aI\rangle /\langle T_{4a}\rangle,
\eea
where $D^{(\text{II})}_2\cong D_2$.
The symmetry breaking pattern is given by
\bea
(\mathbb{Z}_2\times \mathbb{Z}_2)\ltimes D_{4h}\rightarrow (\mathbb{Z}_2\times \mathbb{Z}_2)\ltimes D_2^{\text{(II)}}.
 \label{eq:sym_breaking_LLRR}
\eea

The general magnetic pattern can be figured out by requiring invariance under $H^{(\text{II})}$, which gives
\bea
\vec{S}^{(\text{II})\prime}_j&=&N_2(0,1,0)^T,
\label{eq:order_S_prime_general_II}
\eea
in which $N_2$ is the magnitude of the spin ordering (for derivations, see Appendix \ref{app:determine_orde2}).
Rotating back to the original frame by performing $(U_4)^{-1}$, we obtain
\bea
\vec{S}^{(\text{II})}_{1+4n}&=&N_2 (0,1,0)^T,\nn\\
\vec{S}^{(\text{II})}_{2+4n}&=&N_2 (0,-1,0)^T,\nn\\
\vec{S}^{(\text{II})}_{3+4n}&=&N_2 (0,-1,0)^T,\nn\\
\vec{S}^{(\text{II})}_{4+4n}&=&N_2 (0,1,0)^T.
\label{eq:original_0_II}
\eea
This is a left-left-right-right magnetic order,
in which the spins are along the $y$-direction and reverse their directions every two sites. 
Similar to the spiral phase, the pattern of the magnetic ordering for the other three degenerate ground states in the LLRR phase can be obtained by applying the broken symmetry transformations to Eq. (\ref{eq:original_0_II}). 

From the discussions in this section, we see that the sign of the coupling constant $g$ is crucial to determine the magnetic pattern of the system.
However, a pure symmetry analysis is not able to give the sign of $g$.
The actual magnetic order developed in  the system has to be determined by numerics as will be discussed in Sec. \ref{subsec:DMRG_KH}. 


\subsection{DMRG numerics}
\label{subsec:DMRG_KH}

We present DMRG simulations to provide numerical evidence for LLRR magnetic order in the Kitaev-Heisenberg chain. 
For convenience, we stick to the four-sublattice rotated frame in all numerical calculations in this subsection. 

\subsubsection{Central charge}

\begin{figure*}[htbp]
\begin{center}
\includegraphics[width=16cm]{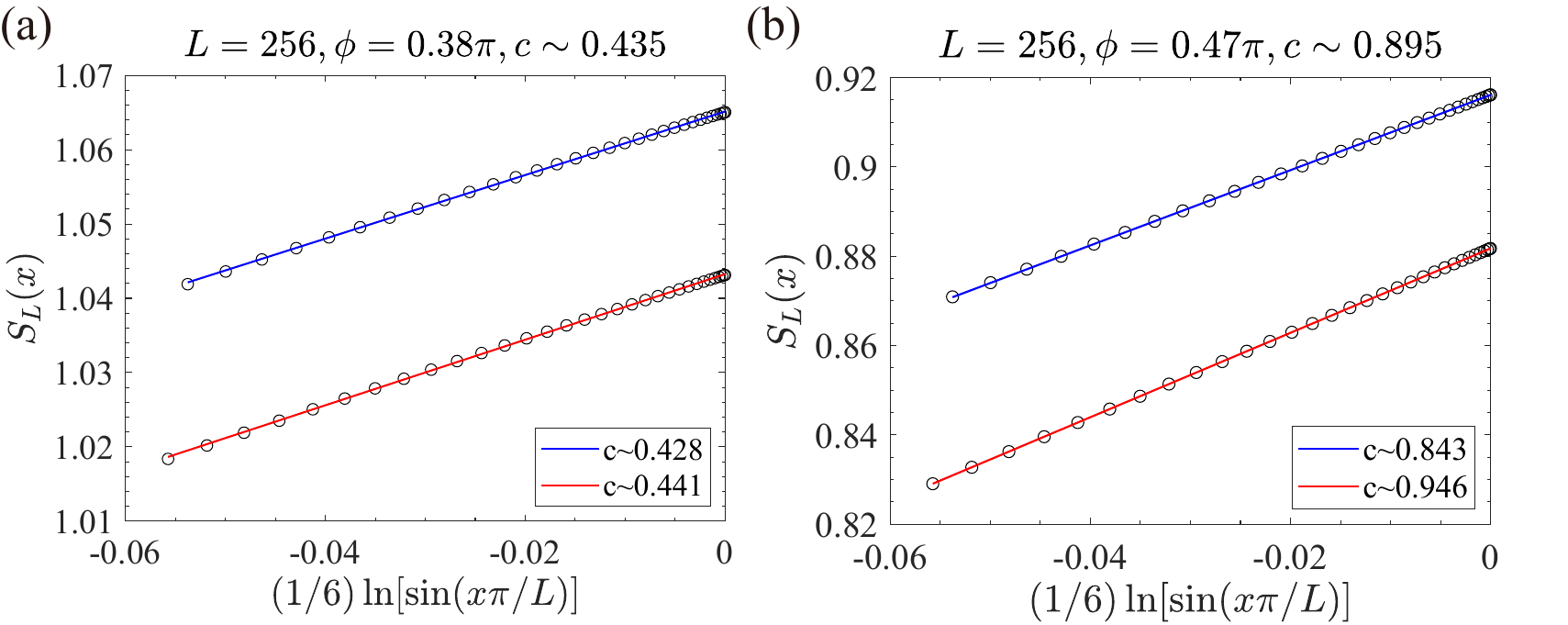}
\caption{Fitting of central charge at (a) $\phi=0.38\pi$ and (b) $\phi=0.47\pi$ in the Kitaev-Heisenberg model using the formula in Eq. (\ref{eq:SL_formula}) for open boundary conditions.
The central charge is taken as the average value of the fitted slopes. 
DMRG numerics are performed on an open chain of $L=256$ sites. 
The bond dimension $m$ and truncation  error $\epsilon$ in DMRG calculations are taken as $m=1000$, $\epsilon=10^{-11}$.
} \label{fig:central_charge_fit}
\end{center}
\end{figure*}

When the low energy physics of the system is described by a critical theory, 
the entanglement entropy $S_L(x)$ for a subregion of length $x$ in a finite chain of length $L$ is predicted by conformal field theory (CFT)  to scale as \cite{Calabrese2009,Ejima2011}
\begin{align}
    S_L(x) = \lambda\frac{c}{3} \ln \left[ \frac{L}{\pi} \sin\left( \frac{\pi x}{L} \right) \right] + \cdots,
    \label{eq:SL_formula}
\end{align}
in which $c$ is the central charge, 
$\lambda=1$ (and $1/2$) for periodic (and open) boundary conditions \cite{Laflorencie2006}, 
and ``$\cdots$" represents subleading terms.  
According to CFT, Luttinger liquids have central charge $c=1$,
whereas the smallest possible central charge is $c=0.5$ which is the value for Ising critical theory \cite{DiFrancesco1997}. 
Hence,  whether a reliable central charge $c$ can be extracted and its fitted value can be used as criteria to tell if the system is critical or not. 

Fig. \ref{fig:central_charge_fit} (a) and (b) show the fitting of the central charge values at two representative points $\phi=0.38\pi$ and $\phi=0.47\pi$, respectively,  using the formula in Eq. (\ref{eq:SL_formula}) for open boundary conditions. 
The entanglement entropy in both Fig. \ref{fig:central_charge_fit} (a) and (b) exhibit an oscillating behavior,
and the values of the central charge are taken as the average of the fitted slopes of the red and blue lines in each figure. 
It is clear from Fig. \ref{fig:central_charge_fit} (a,b) that while the data for $\phi=0.48\pi$ is consistent with a $c=1$ behavior,
the low energy physics of the point of $\phi=0.37\pi$ is not described by Luttinger liquid theory. 

To investigate the behavior of the central charge in the whole parameter region, 
Fig. \ref{fig:central_charge} shows the fitted values of the central charge in the region $\phi\in(\arctan(2),\pi/2)$ (where $J=\cos(\phi)$, $K=-\sin(\phi)$) extracted from DMRG numerical results for entanglement entropy at different system sizes with open boundary conditions, using the formula in Eq. (\ref{eq:SL_formula}) and the same fitting method as the one used  for Fig. \ref{fig:central_charge_fit}. 
It can observed from Fig. \ref{fig:central_charge} that when $\phi<0.4\pi$, the central charge value decreases by increasing system size, 
having a tendency to approach zero  in the limit of infinite sizes.
This is consistent with the analytic prediction in Sec. \ref{sec:LL} on the existence of a magnetically ordered phase when $\phi$ is close to $\arctan(2)$. 
On the other hand, when $\phi>0.4\pi$, the value of the numerically extracted central charge increases by increasing the system size,
tending to approach the value of $1$ in the large size limit, thereby indicating a $c=1$ gapless phase in the region $\phi\in(0.4\pi,0.5\pi)$.
The low energy field theory describing this gapless phase is non-perturbative in nature, which remains unclear and will be left for future studies. 
Also notice that $\phi=0.4\pi$ is special in Fig. \ref{fig:central_charge} in the sense that the value of $c$ remains nearly the same for all the three system sizes. 
Whether the low energy physics at $\phi=0.4\pi$ is described by one of the minimal models in CFT \cite{DiFrancesco1997} remains an open question, and will be left for future more detailed investigations.  

\begin{figure*}[htbp]
\centering
\includegraphics[width=16cm]{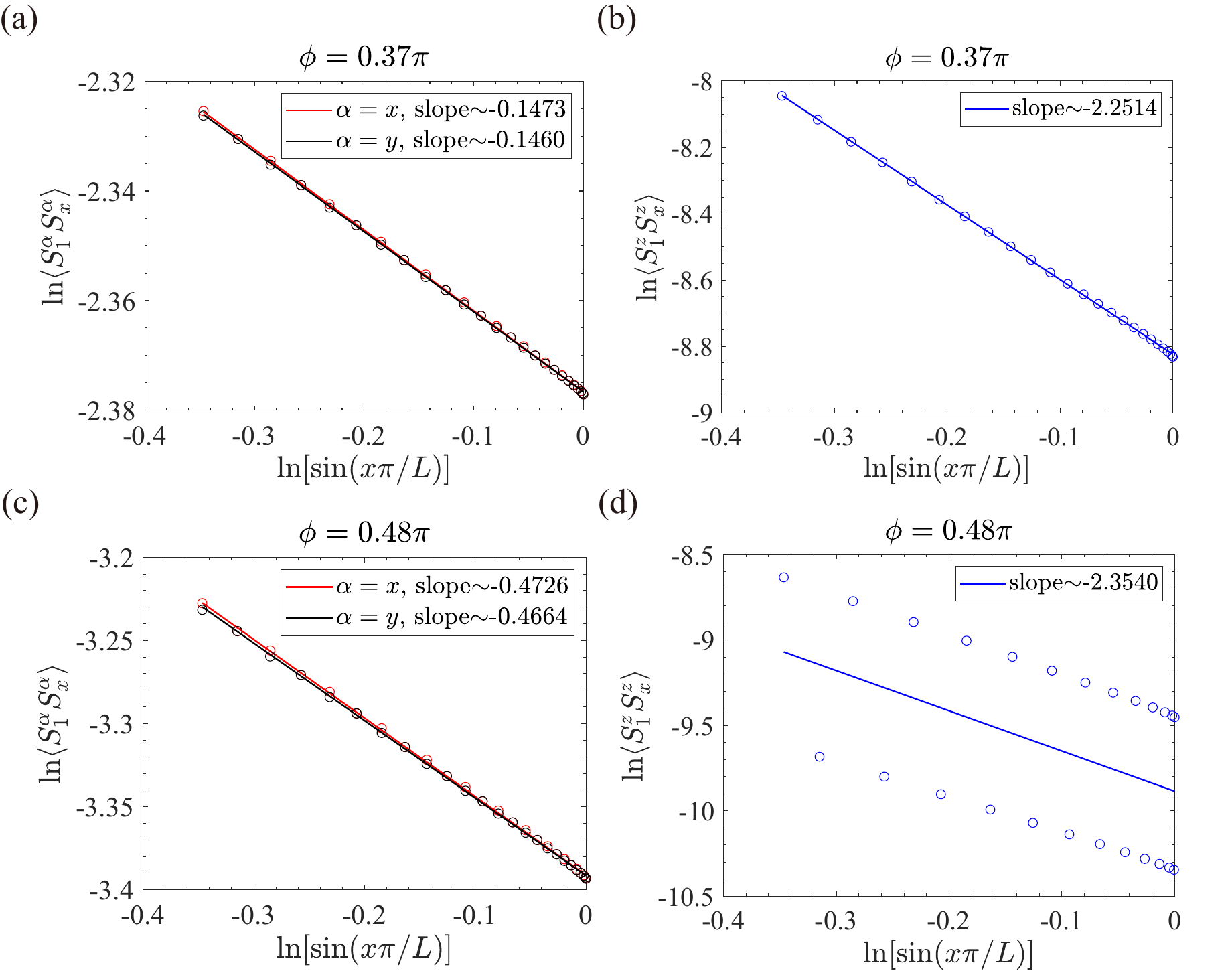}
\caption{Spin correlation functions $\langle S^\alpha_1 S^\alpha_x \rangle$ in the $U_4$ frame versus $\sin(\pi r/L)$ on a log-log scale in the Kitaev-Heisenberg model for (a) $\alpha=x,y$ at $\phi=0.37\pi$, (b) $\alpha=z$ at $\phi=0.37\pi$, (c) $\alpha=x,y$ at $\phi=0.48\pi$, and (d) $\alpha=z$ at $\phi=0.48\pi$.
DMRG numerics are performed on systems of $L=96$ sites using periodic boundary conditions.
The bond dimension $m$ and truncation  error $\epsilon$ in DMRG calculations are taken as $m=1200$, $\epsilon=10^{-10}$.
} 
\label{fig:KH_Corrs}
\end{figure*}

\subsubsection{Correlation functions}

To further distinguish the $\phi<0.4\pi$ region from the  $\phi>0.4\pi$ region, we calculate and compare spin correlation functions $\langle S_1^{\prime\alpha} S_{x}^{\prime\alpha}\rangle$ in the $U_4$ frame ($\alpha=x,y,z$) in the two regions. 
Notice that if the system is described by the Luttinger liquid theory, then the exponent for transverse correlation functions ($\alpha=x,y$) is predicted to be $1/(2\kappa)$, and the exponent for the longitudinal one ($\alpha=z$) is $2\kappa$ \cite{}. 
The product of the values of these two types of exponents is equal to $1$. 

Fig. \ref{fig:KH_Corrs} (a,b) show the transverse and longitudinal correlation functions at $\phi=0.37\pi$, respectively.
As is clear from Fig. \ref{fig:KH_Corrs} (a,b), the exponents of transverse correlation functions at $\phi=0.37\pi$ are nearly equal (approximately $0.147$) and very small,
indicating a slowly decaying correlation behavior.
In particular, the product of transverse and longitudinal exponents is equal to $0.33$, far less than $1$,
indicating a non-Luttinger liquid behavior, consistent with the perturbative Luttinger liquid analysis in Sec. \ref{subsec:Luttinger}.
We note that in the magnetically ordered LLRR phase, the transverse correlation functions should asymptotically approach non-vanishing constant values in the long distance limit,
which seems to be not consistent with the decay behavior in Fig. \ref{fig:KH_Corrs} (a,b).
However, such discrepancy is most possibly a finite size artifact.
Notice that the term $g\cos(4\sqrt{\pi}\theta)$ (see Eq. (\ref{eq:cos4theta})) leading to the LLRR order is a high order effect,  not appearing in the first order perturbation.
Hence the effect of $g\cos(4\sqrt{\pi}\theta)$ only becomes prominent after a long ``time" of RG flow, meaning that the system size has to be very  large. 
For moderate system sizes available in numerical calculations like the one chosen in Fig. \ref{fig:KH_Corrs},
the length scale may be not long enough to observe the correct behavior for magnetic orderings.  

\begin{figure*}[htbp]
\centering
\includegraphics[width=16cm]{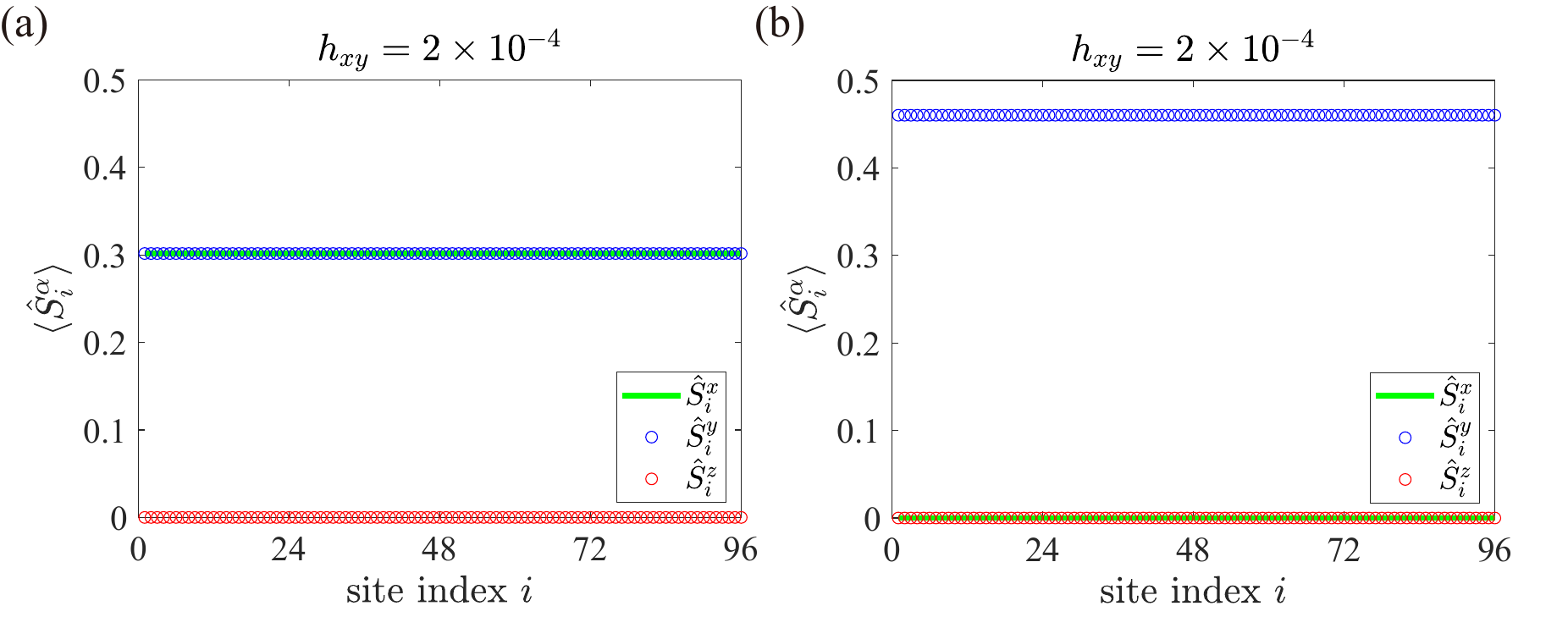}
\caption{Spin expectation values in the Kitaev-Heisenberg model in the $U_4$ frame  throughout the chain at $\phi=0.36\pi$ in the presence of a small field of strength $2\times 10^{-4}$ along (a)  $(1,1,0)$-direction, and (b) $(0,1,0)$-direction. 
DMRG numerics are performed on a system of $L=96$ sites with periodic boundary conditions. 
In (a), the data points for spin expectations values along $x$- and $y$-directions coincide,
whereas in (b), those along $x$- and $z$-directions coincide.
The bond dimension $m$ and truncation  error $\epsilon$ in DMRG calculations are taken as $m=1200$, $\epsilon=10^{-10}$.
} 
\label{fig:KH_spin}
\end{figure*}

Fig. \ref{fig:KH_Corrs} (c,d) show the correlation functions at $\phi=0.48\pi$. 
The product of exponents for the transverse and longitudinal correlations is approximately $1.1$,
which is very close to $1$.
In addition, the central charge value at $\phi=0.48\pi$ obtained in Fig. \ref{fig:central_charge} is close to $1$. 
Both the exponents for correlation functions and the central charge value are consistent with Luttinger liquid theory. 
In addition, exponents for transverse correlations are predicted to be $0.5$ for free fermion systems,
which is close to the numerically obtained values in Fig. \ref{fig:KH_Corrs} (c). 
Therefore, existing DMRG numerics at $\phi=0.48\pi$ are consistent with a low energy theory of a free fermion.
An analytical understanding for the $\phi>0.4\pi$ region is still lacking and will be left for future considerations. 

\subsubsection{Magnetic ordering in the LLRR phase}

Notice that there are two possible types of magnetic orders according to the analysis in Sec. \ref{sec:LL} and Sec. \ref{sec:KHG_chain},
namely, the spiral order and the LLRR order. 
We will study the response of the spin expectation values to a small uniform external magnetic field along certain directions,
where the direction in the $U_4$ frame can be chosen as  $\frac{1}{\sqrt{2}}(\hat{x}+\hat{y})$ and $\hat{y}$ for the spiral and LLRR orders, respectively. 
The added terms $H_{xy}$ and $H_y$ in the  Hamiltonian for the $\frac{1}{\sqrt{2}}(\hat{x}+\hat{y})$ and $\hat{y}$ fields in the $U_4$ frame are given by 
\bea
H_{xy}&=&-\frac{1}{\sqrt{2}}h_{xy}\sum_i (S_i^{\prime x}+S_i^{\prime y}),\nn\\
H_{y}&=&-h_y \sum_i S_i^{\prime y}.
\label{eq:h_field_Ham}
\eea
When there is spontaneous symmetry breaking, it is expected that an infinitesimal field is enough to produce finite spin expectation values 
in the thermodynamic limit. 

We perform DMRG numerics in the four-sublattice rotated frame,
and accordingly, the patterns of spin expectation values of the spiral phase and the LLRR phase in the  $U_4$ frame are given by Eq. (\ref{eq:order_S_prime_general}) and Eq. (\ref{eq:order_S_prime_general_II}), respectively. 
When a uniform $(1,1,0)$-field (or $(0,1,0)$-field)  is applied in the $U_4$ frame, it is expected that the spin pattern in Eq. (\ref{eq:order_S_prime_general}) (or Eq. (\ref{eq:order_S_prime_general_II})) is developed. 
Fig. \ref{fig:KH_spin} (a,b) displays the numerical results of the spin expectation values at $\phi=0.36\pi$ when a $h=2\times 10^{-4}$ field is applied along the $(1,1,0)$- and $(0,1,0)$-directions, respectively. 
As can be clearly seen, the pattern in Fig. \ref{fig:KH_spin} (a) is consistent with Eq. (\ref{eq:order_S_prime_general}),
and the one in Fig. \ref{fig:KH_spin} (b) is consistent with Eq. (\ref{eq:order_S_prime_general_II}).

However, there arises the question which one of the two orders applies to the system. 
Before answering this question in Sec. \ref{subsubsec:energy_field}, we explain why both responses have the correct pattern expected from the corresponding magnetic ordering,
or more precisely, why the system has the response consistent with the spiral order when the actual order is LLRR, and vice versa. 
In particular, notice that because of this subtlety,  the method for calculating the spin expectation values in the presence of a thermodynamically infinitesimal field is not able to distinguish between the two orders. 

Suppose the actual magnetic order of the system is the LLRR order. 
As discussed in Eq. (\ref{eq:sol_theta_I}), the ground states are four-fold degenerate in the thermodynamic limit,
which can be related by each other by the broken symmetry transformations.
Denote $\ket{G_1}$ as the ground state having spin expectation values given in Eq. (\ref{eq:order_S_prime_general}). 
Then the other three ground states $\ket{G_i}$ ($i=2,3,4$) in the four-sublattice rotated frame are given by 
\bea
\ket{G_2}&=&R(\hat{z},\frac{\pi}{2})T_a\ket{G_1}\nn\\
\ket{G_3}&=&[R(\hat{z},\frac{\pi}{2})T_a]^2\ket{G_1}\nn\\
\ket{G_4}&=&[R(\hat{z},\frac{\pi}{2})T_a]^3\ket{G_1},
\label{eq:G1234}
\eea
where the broken symmetry operations are chosen as $[R(\hat{z},\frac{\pi}{2})T_a]^i$ ($i=2,3,4$).
Clearly, the expectation values of spin operators are along $\hat{y}$-, $-\hat{x}$-, $-\hat{y}$-, and $\hat{x}$-directions for $\ket{G_1}$, $\ket{G_2}$, $\ket{G_3}$, and $\ket{G_4}$, respectively. 
As a result, when a small field is applied along the $(1,1,0)$-directions, the energies of the states $\ket{G_1}$ and $\ket{G_4}$ will be lowered by the same amount, whereas the energies of $\ket{G_2}$ and $\ket{G_3}$ will be raised.  
Hence, any linear combination of $\ket{G_1}$ and $\ket{G_4}$ is energetically favored in the thermodynamic limit, though there can be small energy differences in finite systems. 

Now consider the state $\ket{\Psi}$ defined as
\bea
\ket{\Psi}=\frac{1}{\sqrt{2}} (\ket{G_1}+\ket{G_4}).
\label{eq:Psi_def}
\eea
The state $\ket{\Psi}$ is among the degenerate manifold of states when a small field in the $(1,1,0)$-direction is applied. 
On the other hand, as can be easily checked, the spin expectation values in the state $\ket{\Psi}$ have exactly the same pattern as Eq. (\ref{eq:order_S_prime_general}),
which is the pattern supposed to be the one for the  spiral order in the $U_4$ frame. 
This explains why the system has a response to a $(1,1,0)$-field as if it has a sprial order,
although the actual order is LLRR.

\subsubsection{Energy-field relation in the LLRR phase}
\label{subsubsec:energy_field}

\begin{figure}[h]
\begin{center}
\includegraphics[width=8cm]{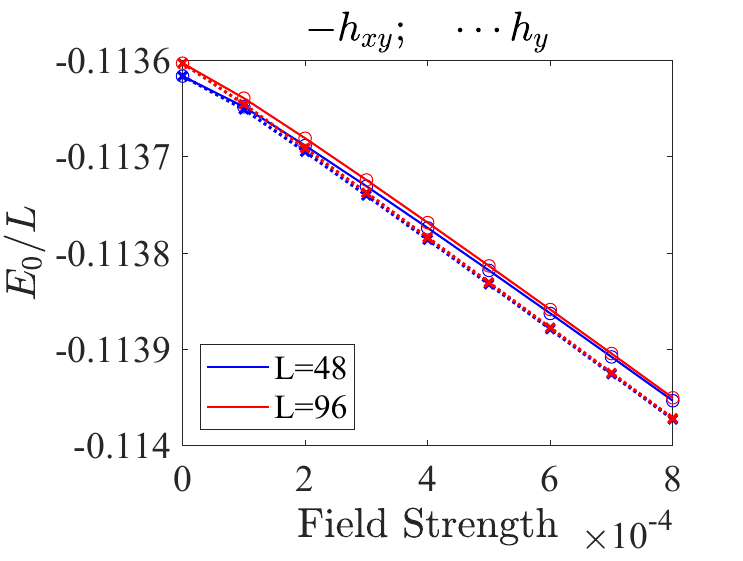}
\caption{Ground state energy per site $E_0/L$ in the Kitaev-Heisenberg model as a function of applied small fields $h_{xy}$ (dashed lines) and $h_y$ (dashed lines),
where $h_{xy}$ and $h_y$ are along $(1,1,0)$- and $(0,1,0)$-directions, respectively.
DMRG numerics are performed on system sizes $L=48$ (blue lines) and $L=96$ (red lines) at $\phi=0.36\pi$ using periodic boundary conditions. 
The bond dimension $m$ and truncation  error $\epsilon$ in DMRG calculations are taken as $m=1200$, $\epsilon=10^{-11}$.
} \label{fig:KH_EnergyField_re}
\end{center}
\end{figure}

To further confirm the existence of magnetic orders in the LLRR phase, 
we calculate ground state energy as a function of applied small fields at $\phi=0.36\pi$.
When the system is magnetically ordered, the energy change is expected to be linear with respect to the applied field in the thermodynamic limit. 
Fig. \ref{fig:KH_EnergyField_re} shows the DMRG numerical results for ground state energy per site $E_0/L$ as a function of the applied small fields along $(1,1,0)$- and $(0,1,0)$-directions for two system sizes $L=48$ and $96$. 
It is clear that the relation between energy change and applied fields are linear in Fig. \ref{fig:KH_EnergyField_re}, confirming the existence of  a magnetic order. 

Furthermore, as a byproduct of Fig. \ref{fig:KH_EnergyField_re},
it can be observed  that under the same field strength,
the energy is lowered more by the $h_y$ field than the $h_{xy}$ field,
which indicates that the actual magnetic order is the LLRR order (given by Eq. (\ref{eq:order_S_prime_general_II}) in the $U_4$ frame), rather than the spiral order (Eq. (\ref{eq:order_S_prime_general}) in the $U_4$ frame). 
In fact, suppose the magnetic order is determined by Eq. (\ref{eq:order_S_prime_general_II}).
Then according to Eq. (\ref{eq:G1234}), we have
\bea
\bra{G_1} \vec{S}_j \ket{G_1} &=&N_2(0,1,0)^T,\nn\\
\bra{G_4} \vec{S}_j \ket{G_4} &=&N_2(1,0,0)^T,
\eea
hence
\bea
\frac{1}{L}\bra{G_1}H_y\ket{G_1}&=&-N_2h_y\nn\\
\frac{1}{L}\bra{\Psi}H_{xy}\ket{\Psi}&=&-\frac{1}{\sqrt{2}}N_2h_{xy},
\eea
where $H_y$ and $H_{xy}$ are defined in Eq. (\ref{eq:h_field_Ham}), and $\ket{\Psi}$ is defined in Eq. (\ref{eq:Psi_def}).
Clearly, if $h_y=h_{xy}=h$, then the energy change $\Delta E_{xy}=\frac{1}{L}\bra{\Psi}H_{xy}\ket{\Psi}$ by the $h_{xy}$ field compared with the zero field case is smaller by a factor of $\frac{1}{\sqrt{2}}$ than the energy change $\Delta E_{y}=\frac{1}{L}\bra{G_1}H_y\ket{G_1}$ by the $h_{y}$ field.
As can be inspected in Fig. \ref{fig:KH_EnergyField_re}, $\Delta E_{y}$ is indeed larger than $\Delta E_{xy}$ when $h_y=h_{xy}$,
consistent with an order along $(0,1,0)$-direction. 
However, the ratio $\Delta E_{xy}/\Delta E_y$ deviates from $\frac{1}{\sqrt{2}}$ significantly.  
This is possibly a finite size effect, since there is still reminiscence of the Luttinger liquid behavior when the system size is not large enough.

\subsubsection{Excitation gap in the $c=1$ phase}

\begin{figure}[h]
\begin{center}
\includegraphics[width=8cm]{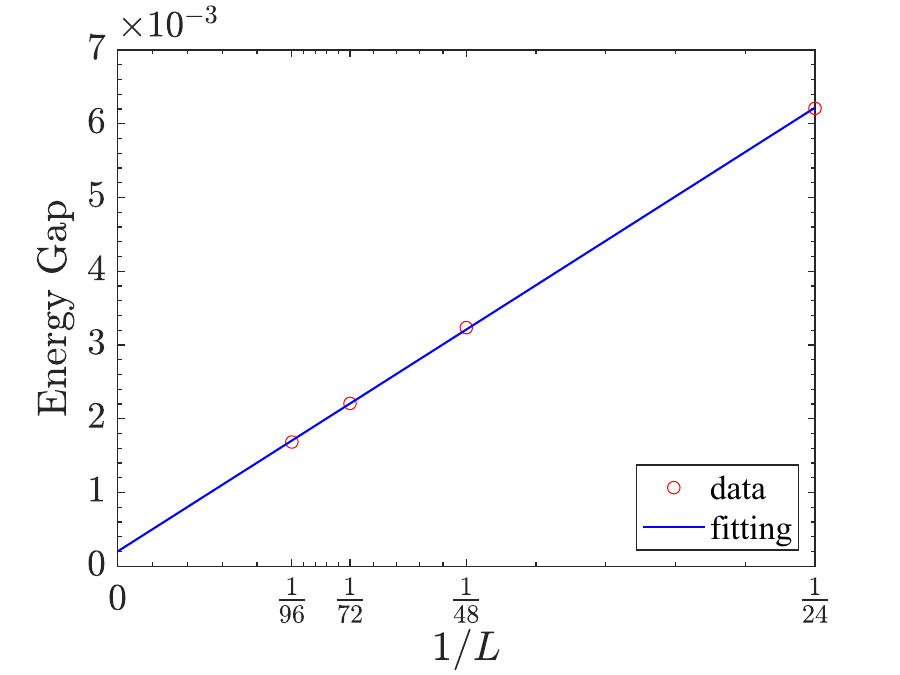}
\caption{Excitation gap vs. $1/L$ in the Kitaev-Heisenberg model at $\phi=0.48\pi$. 
DMRG numerics are performed on system sizes $L=24,48,72,96$ under periodic boundary conditions. 
The bond dimension $m$ and truncation  error $\epsilon$ in DMRG calculations are taken as $m=1200$, $\epsilon=10^{-10}$.
} \label{fig:EnergyGap_L}
\end{center}
\end{figure}

To further confirm the gapless nature of the $\phi>0.4\pi$ region,
we numerically calculate the excitation gap at $\phi=0.48\pi$ to verify the $1/L$ scaling of the gap. 
As shown in Fig. \ref{fig:EnergyGap_L}, 
the excitation gap is clearly linear as a function $1/L$.
The extrapolation of the gap to the $L\rightarrow \infty$ limit gives a gap value of $2\times 10^{-4}$, which is approximately zero,
consistent with a gapless behavior. 

\section{Distorted-LLRR phase in Kitaev-Heisenberg-Gamma chain}
\label{sec:KHG_chain}


In this section, we apply the perturbative Luttinger liquid analysis to the Kitaev-Heisenberg-Gamma chain close to the hidden SU(2) symmetric FM$^\prime$ point, by adding a nonzero Gamma term. 
Because of the equivalence in Eq. (\ref{eq:equiv_Gamma}), we will focus on the $\Gamma>0$ region.

\subsection{Perturbative Luttinger liquid analysis}
\label{subsec:LL_perturbation_KHG}

The strategy is the same as Sec. \ref{subsec:Luttinger}, namely, figuring out the most relevant (in the RG sense) symmetry allowed term in the Luttinger liquid theory. 
The symmetry group $G_1^\prime$ of the Kitaev-Heisenberg-Gamma model in the $U_4$ frame given in Eq. (\ref{eq:group_G1}) is used for this analysis. 
After applying the $V_2$ transformation, the symmetry operations in Eq. (\ref{eq:group_G1}) become
$T,R(\hat{x}^{\prime\prime},\pi)T_aI,R(\hat{z}^{\prime\prime},\frac{\pi}{2})T_a$.
The transformation properties of $\sin(m\sqrt{\pi}\theta),\cos(m\sqrt{\pi}\theta)$ under $T,R(\hat{z}^{\prime\prime},\frac{\pi}{2})T_a$ have been given in Eq. (\ref{eq:theta_Rz}),
and for $R(\hat{x}^{\prime\prime},\pi)T_aI$, the transformation rule is
\bea
R(\hat{x}^{\prime\prime},\pi)T_aI&:& \cos(m\sqrt{\pi}\theta(x))\rightarrow (-)^m\cos(m\sqrt{\pi}\theta(-x)),\nn\\
&&\sin(m\sqrt{\pi}\theta(x))\rightarrow (-)^{m+1}\sin(m\sqrt{\pi}\theta(-x)).\nn\\
\eea
This time, $\sin(m\sqrt{\pi}\theta)$ is again forbidden, since it changes sign under $TR(\hat{x}^{\prime\prime},\pi)T_aI$. 
Because of the symmetries $T$ and $R(\hat{z}^{\prime\prime},-\frac{\pi}{2})T_a$, $m$ has to be a multiple of $4$ in $\cos(m\sqrt{\pi}\theta)$.
Hence the low energy symmetry allowed term with the smallest scaling dimension is still given by Eq. (\ref{eq:cos4theta}).

Similar to the analysis in Sec. \ref{subsec:order}, we need to distinguish between the $g>0$ and $g<0$ cases. 
First consider the $g>0$ case. 
The energy is again minimized at values of $\theta$ in Eq. (\ref{eq:sol_theta_I}). 
However, the unbroken symmetry group is different because of a reduction of the  symmetry group of the model.
It can be checked that the unbroken symmetry group  $H_1^{(\text{I})}$ in the $U_4$ frame (which is a subgroup of $G_1^\prime$ in Eq. (\ref{eq:group_G1}) without the further $V_2$ transformation) is given by
\begin{flalign}
H^{(\text{I})}_1=\langle [R(\hat{z},-\frac{\pi}{2})T_a]^2T,R(\hat{z},-\frac{\pi}{2})T_a\cdot R(\hat{y},\pi)T_aI\rangle,
\label{eq:HI_1}
\end{flalign}
which satisfies
\bea
1\rightarrow \langle T_{4a}\rangle \rightarrow H^{(\text{I})}_1\rightarrow  D^{(\text{I})}_2 \rightarrow 1,
\label{eq:short_exact_d-spiral_unbroken}
\eea
where  $D^{(\text{I})}_2$ is still given by Eq. (\ref{eq:D_2_I}). 
Combining Eq. (\ref{eq:short_exact_1_prime}) with Eq. (\ref{eq:short_exact_d-spiral_unbroken}), it can be observed that the symmetry breaking pattern is
\bea
 D_{4h}\rightarrow D_2^{\text{(I)}}.
\eea

The most general magnetic pattern with an unbroken symmetry group $H^{(\text{I})}_1$ in the $U_4$ frame has been determined in Ref. \onlinecite{Yang2020} as (for derivations, see Appendix \ref{app:determine_orde3})
\begin{eqnarray}
\vec{S}^{\prime (\text{I})}_1&=&(f,f,0)^T,\nn\\
\vec{S}^{\prime (\text{I})}_2&=&(k,k,h)^T,\nn\\
\vec{S}^{\prime (\text{I})}_3&=&(f,f,0)^T,\nn\\
\vec{S}^{\prime (\text{I})}_4&=&(k,k,-h)^T.
\label{eq:Spin_align_d_sp_4rot}
\end{eqnarray}
Rotating back to the original frame, the spin ordering is
\begin{eqnarray}
\vec{S}_1^{(\text{I})}&=&(-f,f,0)^T,\nn\\
\vec{S}_2^{(\text{I})}&=&(-k,-k,h)^T,\nn\\
\vec{S}_3^{(\text{I})}&=&(f,-f,0)^T,\nn\\
\vec{S}_4^{(\text{I})}&=&(k,k,-h)^T.
\label{eq:spin_align_d_sp_orig}
\end{eqnarray}
This is a distorted spiral order, which gains components in the $z$-direction compared with the magnetic ordering in Eq. (\ref{eq:original_0}) for $\Gamma=0$. 
Clearly, Eq. (\ref{eq:original_0}) can be obtained from Eq. (\ref{eq:spin_align_d_sp_orig}) by setting $f=k=\frac{1}{\sqrt{2}}N_1$ and $h=0$.
The pattern of the magnetic ordering for the other three degenerate ground states can be obtained by applying the broken symmetry transformations to Eq. (\ref{eq:spin_align_d_sp_orig}). 

Next  consider the $g<0$ case. 
The energy is again minimized at values of $\theta$ in Eq. (\ref{eq:theta_II}). 
It can be checked that the unbroken symmetry group  $H_1^{(\text{II})}$ is given by
\bea
H^{(\text{II})}_1=\langle [R(\hat{z},-\frac{\pi}{2})T_a]^2T,R(\hat{y},\pi)T_aI\rangle,
\label{eq:HII_1}
\eea
which satisfies
\bea
1\rightarrow \langle T_{4a}\rangle \rightarrow H^{(\text{II})}_1\rightarrow  D^{(\text{II})}_2 \rightarrow 1,
\label{eq:short_exact_d-LLRR_unbroken}
\eea
where  $D^{(\text{II})}_2$ is still given by Eq. (\ref{eq:D_2_I}). 
Combining Eq. (\ref{eq:short_exact_1_prime}) with Eq. (\ref{eq:short_exact_d-LLRR_unbroken}), it can be observed that the symmetry breaking pattern is
\bea
 D_{4h}\rightarrow D_2^{\text{(II)}}.
 \label{eq:sym_breaking_d-LLRR}
\eea

The most general magnetic pattern with an unbroken symmetry group $H^{(\text{II})}_1$ in the $U_4$ frame can be determined as
(for derivations, see Appendix \ref{app:determine_orde4})
\begin{eqnarray}
\vec{S}^{\prime (\text{II})}_1&=&(a,b,-c)^T,\nn\\
\vec{S}^{\prime (\text{II})}_2&=&(-a,b,-c)^T,\nn\\
\vec{S}^{\prime (\text{II})}_3&=&(a,b,c)^T,\nn\\
\vec{S}^{\prime (\text{II})}_4&=&(-a,b,c)^T.
\label{eq:Spin_align_d_sp_4rot_II}
\end{eqnarray}
Rotating back to the original frame, the spin ordering is
\begin{eqnarray}
\vec{S}_1^{(\text{II})}&=&(-a,b,c)^T,\nn\\
\vec{S}_2^{(\text{II})}&=&(a,-b,-c)^T,\nn\\
\vec{S}_3^{(\text{II})}&=&(a,-b,-c)^T,\nn\\
\vec{S}_4^{(\text{II})}&=&(-a,b,c)^T.
\label{eq:spin_align_d_sp_orig_II}
\end{eqnarray}
This is a distorted left-left-right-right  magnetic order. 
Clearly, Eq. (\ref{eq:original_0_II}) can be obtained from Eq. (\ref{eq:spin_align_d_sp_orig_II}) by setting $b=\frac{1}{\sqrt{2}}N_2$, $c=0$.
The pattern of the magnetic ordering for the other three degenerate ground states in this case can be obtained by applying the broken symmetry transformations to Eq. (\ref{eq:spin_align_d_sp_orig_II}). 

Finally we note that the distorted-LLRR phase in the Kitaev-Heisenberg-Gamma model and the LLRR phase in the Kitaev-Heisenberg model belong to the same phase, since the broken symmetries in both cases are the same, as can be inspected from Eq. (\ref{eq:sym_breaking_LLRR}) and Eq. (\ref{eq:sym_breaking_d-LLRR}).
Similar conclusion holds  for the distorted-spiral and spiral phases. 

\subsection{DMRG numerics}

In this subsection, we provide DMRG numerical evidence for distorted-LLRR order in the Kitaev-Heisenberg-Gamma chain by including a small Gamma term. 
The following parametrization will be used for $K,J,\Gamma$ in this subsection
 \begin{eqnarray}
 J&=&\cos(\theta)\nn\\ 
 K&=&\sin(\theta)\cos(\phi)\nn\\
 \Gamma&=&\sin(\theta)\sin(\phi),
\label{eq:parametrization2}
\end{eqnarray}
in which $\theta\in[0,\pi]$, $\phi\in[0,2\pi]$.
Notice that in terms of Eq. (\ref{eq:parametrization2}),
the point $(\theta,\pi)$ corresponds to the Kitaev-Heisenberg model at angle $\phi$ using the parametrization in Eq. (\ref{eq:parametrization}).
We will take a representative point $(\theta=0.37\pi,\phi=0.99\pi)$ in DMRG numerical calculations in this subsection,
which has a small nonzero value of $\Gamma$ since $\phi$ is close to $\pi$.
We stick to the four-sublattice rotated frame in all numerical calculations in this subsection. 




\subsubsection{Magnetic ordering}

\begin{figure*}[htbp]
\begin{center}
\includegraphics[width=16cm]{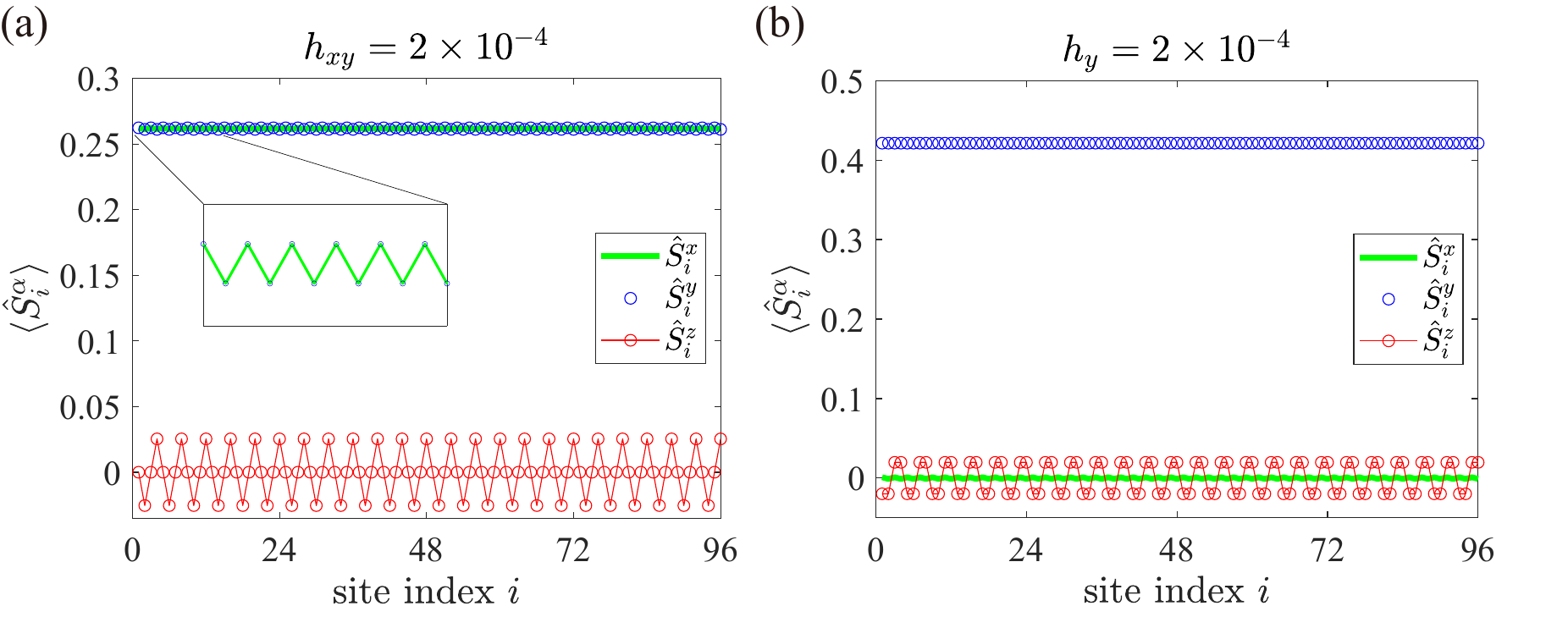}
\caption{
Spin expectation values in the Kitaev-Heisenberg-Gamma model throughout the chain at $(\theta=0.37\pi,\phi=0.99\pi)$ in the presence of a small field of strength $2\times 10^{-4}$ along (a)  $(1,1,0)$-direction, and (b) $(0,1,0)$-direction. 
DMRG numerics are performed on a system of $L=96$ sites with periodic boundary conditions. 
In (a), the data points for expectation values along $x$ and $y$ spin directions coincide;
and the inset figure is a zoom-in view of the spin expectation values along $x$- and $y$-directions. 
The bond dimension $m$ and truncation  error $\epsilon$ in DMRG calculations are taken as $m=1200$, $\epsilon=10^{-10}$.
} \label{fig:KHG_spin}
\end{center}
\end{figure*}

In the Kitaev-Heisenberg-Gamma model,
there are again two types of magnetic orders according to the analysis in Sec. \ref{subsec:LL_perturbation_KHG},
similar to the Kitaev-Heisenberg case. 
To test the magnetic order, we numerically calculate spin expectation values by applying small magnetic fields along the $(1,1,0)$- and $(0,1,0)$-directions.

DMRG numerics are performed in the $U_4$ frame at $(\theta=0.37\pi,\phi=0.99\pi)$ for small fields $h_{xy}=2\times 10^{-4}$ and $h_{y}=2\times 10^{-4}$ along $(1,1,0)$- and $(0,1,0)$-directions, respectively,
and the results for spin expectation values are shown in Fig. \ref{fig:KHG_spin}.
As can be observed, the patterns of spin expectation values in Fig. \ref{fig:KHG_spin} (a) for $h_{xy}$ and Fig. \ref{fig:KHG_spin} (b) for $h_y$
are consistent with the predictions in the $U_4$ frames for the distorted-spiral order in Eq. (\ref{eq:Spin_align_d_sp_4rot}) and the distorted-LLRR  order in Eq. (\ref{eq:Spin_align_d_sp_4rot_II}), respectively. 
Similar to the case of Kitaev-Heisenberg model, there is the question which one is the actual order in the system,
which will be answered shortly by the study of energy-field relation  in Sec. \ref{subsubsec:energy-field_KHG}.

\subsubsection{Energy-field relation}
\label{subsubsec:energy-field_KHG}

\begin{figure}[h]
\begin{center}
\includegraphics[width=8cm]{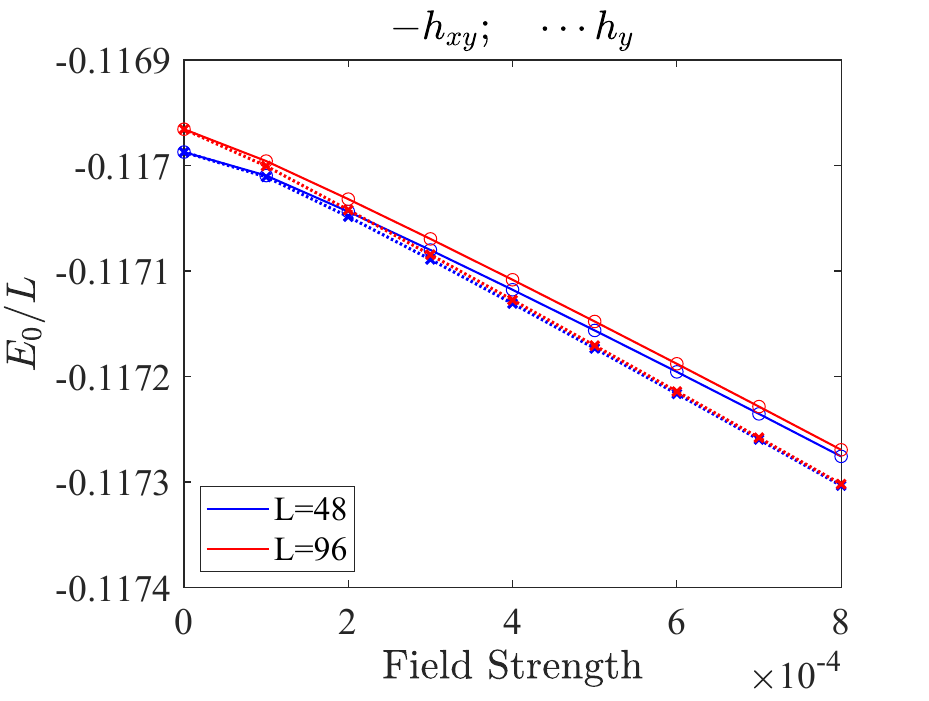}
\caption{
Ground state energy per site $E_0/L$ in the Kitaev-Heisenberg-Gamma model as a function of applied small fields $h_{xy}$ (dashed lines) and $h_y$ (dashed lines),
where $h_{xy}$ and $h_y$ are along $(1,1,0)$- and $(0,1,0)$-directions, respectively.
DMRG numerics are performed on system sizes $L=48$ (blue lines) and $L=96$ (red lines) at $(\theta=0.37\pi,\phi=0.99\pi)$ using periodic boundary conditions. 
The bond dimension $m$ and truncation  error $\epsilon$ in DMRG calculations are taken as $m=1200$, $\epsilon=10^{-10}$.
} \label{fig:KHG_EnergyField}
\end{center}
\end{figure}

Fig. \ref{fig:KHG_EnergyField} shows DMRG numerical results for ground state energy per site $E_0/L$  at  $(\theta=0.36\pi,\phi=0.99\pi)$ as a function of the applied small fields along $(1,1,0)$- and $(0,1,0)$-directions in the $U_4$ frame for two system sizes $L=48$ and $96$. 
It is clear that the relation between energy change and applied fields are linear in Fig. \ref{fig:KHG_EnergyField}, confirming the existence of  a magnetic order. 
As can be inspected in Fig. \ref{fig:KHG_EnergyField}, the energy changes $\Delta E_{y}$  is indeed larger than $\Delta E_{xy}$ when $h_y=h_{xy}$,
consistent with the distorted-LLRR order rather than the distorted-spiral order
according to a similar analysis as Sec. \ref{subsubsec:energy_field}. 

\section{Other phases in Kitaev-Heisenberg-Gamma chain}
\label{sec:other_phase}

In this section, we make comments on other phases in the Kitaev-Heisenberg-Gamma model in Fig. \ref{fig:phase},
especially the FM phase. 
The antiferromagnetic phase (denoted as ``AFM"), the Luttinger liquid phase (denoted as ``LL"),
and the ferromagnetic phase (denoted as ``FM") have been investigated in details both analytically and numerically in Ref. \onlinecite{Yang2020}.
Interestingly, the FM phase in the $J<0$ region can also be derived in the framework of perturbative Luttinger liquid analysis in a similar manner as Sec. \ref{subsec:LL_perturbation_KHG}, which will be shown below. 

\subsection{The FM phase}

For the Kitaev-Heisenberg model, the region between $-K$ and $-J$ on the horizontal axis in Fig. \ref{fig:phase} is unitarily equivalent to the region between $-K$ and FM$^\prime$ under the $U_4$ transformation.
Hence, in the neighborhood of $-J$,
the system is in an ordered phase,
and the magnetic pattern in the original frame in this case is exactly the same as the pattern in the $U_4$ frame in the vicinity of FM$^\prime$. 
Therefore, in the region $[\phi_{c2},\pi]$ on the horizontal axis in Fig. \ref{fig:phase} where $\phi_{c2}$ is dual to $\phi_{c1}$ under the equivalent relation in Eq. (\ref{eq:equiv_KJ}),
we conclude that for $g>0$,
the magnetic ordering is along $\frac{1}{\sqrt{2}}(\pm1,\pm1,0)$-directions (see Eq. (\ref{eq:order_S_prime_general}) and the other three degenerate solutions);
and for $g<0$, 
the magnetic ordering is along $\pm \hat{x}$- and $\pm \hat{y}$-directions (see Eq. (\ref{eq:order_S_prime_general_II}) and the other three degenerate solutions).

Next we consider the Kitaev-Heisenberg-Gamma model with a nonzero Gamma term.
This time, the analysis in Sec. \ref{subsec:Luttinger} cannot be applied, since the perturbative Luttinger liquid analysis should now be applied in the original frame, not the four-sublattice rotated frame. 
Performing $V_2$ in Eq. (\ref{eq:V_2}) to the Hamiltonian $H_1$ in Eq. (\ref{eq:Ham1}) in the original frame, the Hamiltonian $\tilde{H}_1=V_2 H_1 (V_2)^{-1}$ becomes
\begin{flalign}
\tilde{H}_1=\tilde{H}_{XXZ}+H_K+H_\Gamma,
\label{eq:Ham1}
\end{flalign}
in which 
\bea
\tilde{H}_{XXZ}&=& |J+\frac{1}{2}K| \sum_i[ \tilde{S}_i^{ x}\tilde{S}_{i+1}^{ x}+\tilde{S}_i^{ y}\tilde{S}_{i+1}^{ y}+\Delta \tilde{S}_i^{ z}\tilde{S}_{i+1}^{ z}]\nn\\
H_K&=&\frac{1}{2}K \sum_i (-)^{i} (\tilde{S}_i^{ x}\tilde{S}_{i+1}^{ x}-\tilde{S}_i^{ y}\tilde{S}_{i+1}^{ y})\nn\\
H_\Gamma&=&\Gamma\sum_{<ij>\in\gamma\,\text{bond}} (\tilde{S}_i^\alpha \tilde{S}_j^\beta-\tilde{S}_i^\beta \tilde{S}_j^\alpha),
\eea
where $\tilde{S}_i$ represents the spin operators after the $V_2$ transformation,
and 
\bea
\Delta=-\frac{J}{J+K/2}.
\eea
Notice that both $J$ and $K$ are negative, hence $\Delta<0$ and $|\Delta|<1$.
This means that the $\tilde{H}_{XXZ}$ term represents an easy-plane XXZ model,
whose low energy physics is described by the Luttinger liquid Hamiltonian.  
We will take $\tilde{H}_{XXZ}$ as the unperturbed system and treat $H_K$ and $H_\Gamma$ as perturbations.
Similar to Sec. \ref{subsec:Luttinger},
$\tilde{H}_{XXZ}$ has a diverging Luttinger parameter when $K\rightarrow 0$.
Therefore, the system is fragile to perturbations of the form $e^{im\sqrt{\pi}\cos(\theta)}$ ($m\in \mathbb{Z}$).

The symmetries in the original frame are given in Eq. (\ref{eq:GN}),
which become the following operations after the $V_2$ transformation
\bea
V_2T(V_2)^{-1} &= &T, \nn\\
V_2T_a I(V_2)^{-1}&= & R(\hat{z},\pi) T_a I, \nn\\
V_2R(\hat{n}_1,\pi)T_a(V_2)^{-1}&=&R(\hat{y},\pi) R(\hat{z},\frac{\pi}{2})T_a.
\label{eq:sym_FM}
\eea
Under the symmetry operations in Eq. (\ref{eq:sym_FM}), the vertex operators $e^{im\sqrt{\pi}\cos(\theta)}$ transform as
\bea
T&:& \cos(m\sqrt{\pi}\theta)\rightarrow (-)^m \cos(m\sqrt{\pi}\theta),\nn\\
&&\sin(m\sqrt{\pi}\theta)\rightarrow (-)^m \sin(m\sqrt{\pi}\theta),\nn\\
R(\hat{z},\pi) T_a I&:&\cos(m\sqrt{\pi}\theta(x))\rightarrow \cos(m\sqrt{\pi}\theta(-x)),\nn\\
&&\sin(m\sqrt{\pi}\theta(x))\rightarrow \sin(m\sqrt{\pi}\theta(-x)),\nn\\
R(\hat{y},\pi) R(\hat{z},\frac{\pi}{2})T_a&:& \cos(m\sqrt{\pi}\theta)\rightarrow  \cos(m\sqrt{\pi}\theta+\frac{m\pi}{2}),\nn\\
&&\sin(m\sqrt{\pi}\theta)\rightarrow -\sin(m\sqrt{\pi}\theta-\frac{m\pi}{2}).\nn\\
\label{eq:Vertex_FM}
\eea
As can be seen from Eq. (\ref{eq:Vertex_FM}), apart from the $\cos(4\sqrt{\pi}\theta)$ term,
the $\sin(2\sqrt{\pi}\theta)$ term is also allowed by symmetries,
which has a smaller scaling dimension than $\cos(4\sqrt{\pi}\theta)$. 
Keeping only the most relevant term, the  low energy field theory becomes
\bea
\mathcal{H}=H_{LL}+u\int dx \sin(2\sqrt{\pi}\theta),
\label{eq:low_FM}
\eea
in which $H_{LL}$ is the Luttinger liquid Hamiltonian describing the low energy physics of $\tilde{H}_{XXZ}$.
To determine the magnetic pattern, we need to distinguish between the two cases $u>0$ and $u<0$.

First consider the $u>0$ case. 
Then the $u$ term in Eq. (\ref{eq:low_FM}) is minimized at
\bea
\theta^{(\text{I})}_n = (n-\frac{1}{4})\sqrt{\pi},~0\leq n\leq 1,
\eea
which are two-fold degenerate.
Take $n=0$ as an example.
The spin ordering can be determined from the bosonization formulas in Eq. (\ref{eq:abelian_bosonize}) as
\bea
\vec{\tilde{S}}^{(\text{I})}_j&=&(-)^j\frac{1}{\sqrt{2}}(1,-1,0)^T.
\label{eq:order_FM_S_V2_I}
\eea
Performing $(V_2)^{-1}$, the magnetic pattern in the original frame is 
\bea
\vec{S}^{(\text{I})}_j&=&\frac{1}{\sqrt{2}}(1,-1,0)^T.
\label{eq:order_FM_S_I}
\eea
The unbroken symmetry group can be determined as
\bea
H^{(\text{I})}_2=\langle T_aI,R(\hat{n}_1,\pi)T_a\rangle,
\label{eq:H_2^II}
\eea
and the most general from of the magnetic pattern invariant under $H^{(\text{I})}$ can be determined as
 (for derivations, see Appendix \ref{app:determine_FM_I})
\bea
\vec{S}^{(\text{I})}_j=N_3(1,-1,0)^T,
\label{eq:spin_FM_general_I}
\eea
where $N_3$ is the magnitude of the spin ordering. 
Clearly, Eq. (\ref{eq:spin_FM_general_I}) is an FM order along $\frac{1}{2}(1,-1,0)^T$-direction.

Next consider the $u<0$ case. 
Then the $u$ term in Eq. (\ref{eq:low_FM}) is minimized at
\bea
\theta^{(\text{II})}_n = (n+\frac{1}{4})\sqrt{\pi},~0\leq n\leq 1,
\eea
which are again two-fold degenerate.
Take $n=0$ as an example.
The spin ordering can be determined from the bosonization formulas in Eq. (\ref{eq:abelian_bosonize}) as
\bea
\vec{\tilde{S}}^{(\text{II})}_j&=&(-)^j\frac{1}{\sqrt{2}}(1,1,0)^T.
\label{eq:order_FM_S_V2}
\eea
Performing $(V_2)^{-1}$, the magnetic pattern in the original frame is 
\bea
\vec{S}^{(\text{II})}_j&=&\frac{1}{\sqrt{2}}(1,1,0)^T.
\label{eq:order_FM_S}
\eea
The unbroken symmetry group can be determined as
\bea
H^{(\text{II})}=\langle T_aI,TR(\hat{n}_1,\pi)T_a\rangle.
\label{eq:H_2^II_2}
\eea
Notice that the magnetic ordering in Eq. (\ref{eq:order_FM_S}) only represents a special configuration consistent with the unbroken symmetry group $H^{(\text{II})}$. 
In fact, the most general from of the magnetic pattern invariant under $H^{(\text{II})}$ can be determined as (for derivations, see Appendix \ref{app:determine_FM_II})
\bea
\vec{S}^{(\text{II})}_j=(a,a,b)^T,
\label{eq:spin_FM_general}
\eea
which is an FM order along $\frac{1}{\sqrt{2a^2+b^2}}(a,a,b)^T$-direction.

We note that the FM phase has been studied in Ref. \onlinecite{Yang2020},
where DMRG numerics give the FM pattern in Eq. (\ref{eq:spin_FM_general}),
indicating a negative $u$. 
The analysis in this section provides yet another explanation to the FM phase in the $J<0$ phase, complementing the several explanations for the origin of the FM phase in Ref. \onlinecite{Yang2020}.

\subsection{The KSL phase}

We note that it remains unclear whether the $c=1$ phase in the Kitaev-Heisenberg model extends to a Kitaev spin liquid phase when $\Gamma$ is nonzero in Kitaev-Heisenberg-Gamma model as shown in Fig. \ref{fig:phase} (denoted as ``KSL" in the figure, which is ``Kitaev spin liquid for short"), which will be left for future investigations.

\section{Weakly coupled chains: 2D stripy order}
\label{sec:stripy}

Finally, we discuss how to obtain the 2D stripy order in the spin-1/2 Kitaev-Heisenberg model on the honeycomb lattice using a coupled-chain method based on the Luttinger liquid analysis. 
The generalization to the 2D spin-1/2 Kitaev-Heisenberg-Gamma model with a small nonzero Gamma interaction is straightforward. 

\begin{figure*}
\begin{center}
\includegraphics[width=15cm]{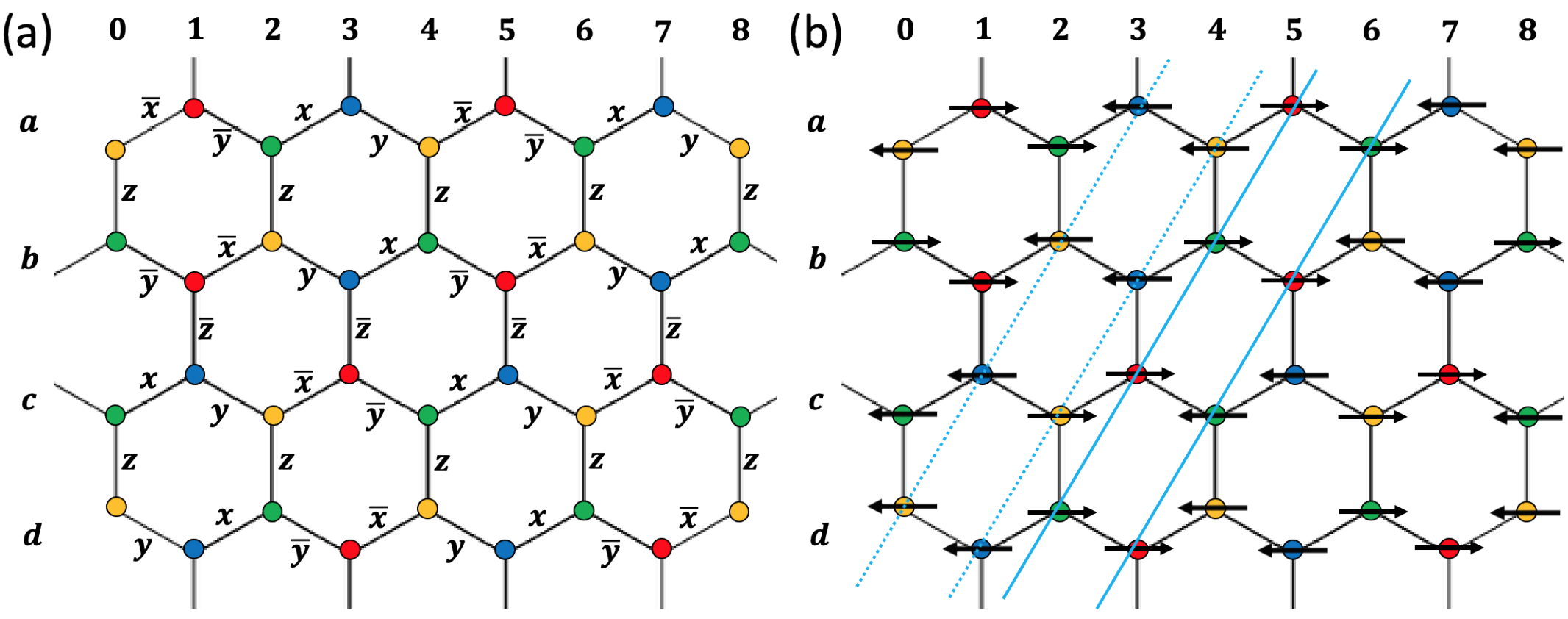}
\caption{(a) Four-sublattice rotation on the honeycomb lattice, and (b) stripy order in the original frame. 
In (b), the arrows represent the spin expectation values in the original frame which are along $\pm \hat{y}$-directions;
the solid and dashed lines connect the sites which have the same spin expectation values. 
} \label{fig:2D_U4}
\end{center}
\end{figure*}

The four-sublattice rotation in the 2D case is shown in Fig. \ref{fig:2D_U4}, in which the sites marked with green, blue, yellow, and red colors are applied by $R(\hat{y},\pi)$, $R(\hat{z},\pi)$, $R(\hat{x},\pi)$, and identity operations, respectively.
After aplying the four-sublattice rotation on the honeycomb lattice, the Hamiltonian of the Kitaev-Heisenberg model on bond $\gamma$ in Fig. \ref{fig:2D_U4} acquires the form 
\bea
H^\prime_{ij}=(K_\gamma+2J_\gamma)S_i^{\prime \gamma} S_j^{\prime\gamma}-J_\gamma \vec{S}_i^\prime\cdot \vec{S}_j^\prime,
\label{eq:2D_Hamiltonian}
\eea
in which $K_x=K_y=K$, $J_x=J_y=J$, $\Gamma_x=\Gamma_y=\Gamma$, $K_z=\alpha_0 K$, $J_z=\alpha_0 J$, $\Gamma_z=\alpha_0 \Gamma$.
We will consider the limit $\alpha_0\ll 1$, i.e., the chains are weakly coupled. 

Let's consider row $c$ in in Fig. \ref{fig:2D_U4}. 
When $\alpha_0=0$, the low energy theory of row $c$ is described by the Luttinger liquid Hamiltonian 
\begin{flalign}
H_{cc}=\frac{v}{2} \int dx [\kappa^{-1} (\nabla \varphi)^2 +\kappa (\nabla \theta)^2]+g\int dx \cos(4\sqrt{\pi}\theta),
\end{flalign}
in which $g<0$ as discussed in Sec. \ref{subsec:order}.  
By performing a mean field treatment, the Hamiltonian of row $c$ becomes 
\bea
H^\prime_c=H^\prime_{cc}+H^\prime_{cb}+H^\prime_{cd},
\label{eq:H_c_prime}
\eea
in which the interaction Hamiltonians $H_{cb}$ and $H_{cd}$ between row $c$ and rows $b,d$ are given by
\bea
H_{cb}^\prime&=&\alpha_0\sum_n \big[ (K+2J) S_{c,2+2n}^{\prime z} \langle S^{\prime z}_{b,2+2n}\rangle\nn\\
&& - J\vec{S}^\prime_{c,2+2n} \cdot\langle \vec{S}^\prime_{b,2+2n}\rangle
\big]
\eea
and
\bea
H^\prime_{cd}&=&\alpha_0\sum_n \big[ (K+2J) S_{c,1+2n}^{\prime z} \langle S^{\prime z}_{d,1+2n}\rangle\nn\\
&& - J\vec{S}^\prime_{c,1+2n}\cdot \langle \vec{S}^\prime_{d,1+2n}\rangle
\big]. 
\eea

Performing $(V_2)^{-1}$ to Eq. (\ref{eq:abelian_bosonize}), we obtain
\bea
S_c^{\prime y} =C \sin(\sqrt{\pi} \theta),
\eea
in which $C=\text{const.}\frac{1}{\sqrt{a}}$.
We consider the following ansatz
\bea
\vec{S}^{\prime}_{c,j}=\vec{S}^{\prime}_{b,j}=\vec{S}^{\prime}_{d,j}=C \langle \sin(\sqrt{\pi} \theta)\rangle (0,1,0)^T.
\label{eq:ansatz}
\eea
Plugging Eq. (\ref{eq:ansatz}) into Eq. (\ref{eq:H_c_prime}), the mean field Hamiltonian can be simplified as 
\bea
H^\prime_c&=&\frac{v}{2} \int dx [\kappa^{-1} (\nabla \varphi)^2 +\kappa (\nabla \theta)^2]+g\int dx \cos(4\sqrt{\pi}\theta)\nn\\
&&-\alpha_0 C^2 \langle \sin(\sqrt{\pi} \theta) \rangle  \int dx \sin(\sqrt{\pi} \theta). 
\label{eq:H_c_prime_3}
\eea
Notice that $\theta=\frac{1}{2}\sqrt{\pi}$ (corresponding to $n=1$ in Eq. (\ref{eq:theta_II})) can minimize both the $g$ and $\alpha_0$ terms in Eq. (\ref{eq:H_c_prime_3}). 
Hence, in the $U_4$ frame, the system has an FM order 
\bea
\langle \vec{S}^\prime_{i,j}\rangle\equiv A(0,1,0)^T,
\eea
in which $A=C \langle \sin(\sqrt{\pi} \theta)\rangle$. 

Applying $(U_4)^{-1}$ and rotating back to the original frame, the pattern of spin expectation values is shown in Fig. \ref{fig:2D_U4} (b),
which is consistent with the  results of classical analysis and  exact diagonalization in Ref. \onlinecite{Rau2014}. 
Notice that the pattern in Fig. \ref{fig:2D_U4} (b) is a 2D stripy order.
Hence, we have obtained an explanation of the 2D stripy phase based on a quasi-1D Luttinger liquid analysis.

\section{Summary}
\label{sec:summary}

In summary, we have studied the phase diagram of the spin-1/2 Kitaev-Heisenberg chain in the $(K<0,J>0)$ region, based on a combination of  perturbative Luttinger liquid analysis and DMRG numerics. 
Close to the hidden SU(2) symmetric FM point, an ordered phase is identified having a left-left-right-right magnetic order,
whereas close to the FM Kitaev point, a gapless phase with central charge value $c=1$ is found. 
The Luttinger liquid analysis is extended to the Kitaev-Heisenberg-Gamma model as well as the parameter region of $(K<0,J<0)$.
Furthermore, the 2D stripy phase is recovered based on the 1D left-left-right-right order using a coupled-chain method,
thereby providing a quasi-1D explanation for the stripy order on the honeycomb lattice.

\begin{acknowledgments}

W.Y. is supported by the National Natural Science Foundation of
China (Grants No. 12474476). 
C.X. acknowledges the supports from MOST grant No. 2022YFA1403900, NSFC No. 12104451, NSFC No. 11920101005, and funds from Strategic Priority Research Program of CAS No. XDB28000000. 
A.N. acknowledges computational resources and services provided by Compute Canada and Advanced Research Computing at the University of
British Columbia. A.N. acknowledges support from the Max Planck-UBC-UTokyo Center for Quantum Materials and the Canada First Research Excellence Fund
(CFREF) Quantum Materials and Future Technologies Program of the Stewart Blusson Quantum Matter Institute (SBQMI).
The DMRG calculations in this work were performed using the software package ITensor in  Ref. \onlinecite{ITensor}.

\end{acknowledgments}

\appendix

\begin{widetext}
\section{Symmetry transformation properties of the bosonization fields}
\label{app:symmetry_transform}

In this appendix, we derive the transformation properties of the bosonization fields $\theta,\varphi$ in Eq. (\ref{eq:LL_liquid}) under symmetry transformations. 
Recall how the Luttinger liquid Hamiltonian $H_{LL}$ is obtained.
First, the spin operators are converted into spinless fermion operators under Jordan-Wigner transformation.
Then the spinless fermion is rewritten in terms of the bosonic fields $\theta,\varphi$ using the standard bosonization formalism. 
A noninteracting fermion model is obtained for the spin-1/2 XX chain, having $\kappa=1$ in Eq. (\ref{eq:LL_liquid}).
When the $S^zS^z$ interaction is introduced, the Luttinger parameter $\kappa$ will deviate from $1$.
Therefore, the strategy of obtaining the transformation properties of  $\theta,\varphi$ is to first derive the transformations of the spinless fermion fields using  the known transformation properties of the spin operators, and then obtain how $\theta,\varphi$ transform in the framework of bosonizing spinless fermions. 

The Jordan-Wigner transformation defines the spinless fermion in terms of the spin operators, given by
\bea
c_j^\dagger&=& S_j^{+}\Pi_{k<j}(-2S_k^z),
\eea
which leads to
\bea
c_j&=& S_j^{-}\Pi_{k<j}(-2S_k^z),\nn\\
c_j^\dagger c_j&=&S_j^z+\frac{1}{2}.
\eea
Using the following transformation properties of the spin operators,
\bea
T_a&:& S_j^+\rightarrow S_{j+1}^+,~S_j^z\rightarrow S_j^z,\nn\\
I&:&S_j^+\rightarrow S_{-j}^+,~S_j^z\rightarrow S_{-j}^z,\nn\\
T&:&S_j^+\rightarrow -S_j^-,~S_j^z\rightarrow -S_j^z,\nn\\
R(\hat{z},\beta)&:& S_j^+\rightarrow e^{i\beta }S_j^+,~S_j^z\rightarrow S_j^z,\nn\\
R(\hat{y},\pi)&:& S_j^+\rightarrow -S_j^-,~S_j^z\rightarrow -S_j^z,
\label{eq:transform_spin}
\eea
the transformations of the spinless fermion can be obtained as
\bea
T_a&:&c_j^\dagger\rightarrow  c_{j+1}^\dagger, \nn\\
I&:&c_j^\dagger\rightarrow c_{-j}^\dagger\cdot\Pi_{k\neq -j}(1-2c_k^\dagger c_k)=c^\dagger_{-j}e^{i\pi S_T^z},\nn\\
T&:&c_j^\dagger\rightarrow(-)^\# (-)^{j}c_j,\nn\\
R(\hat{z},\beta)&:&c_j^\dagger\rightarrow e^{i\beta }c_j^\dagger,\nn\\
R(\hat{y},\pi)&:&c_j^\dagger\rightarrow (-)^\#(-)^jc_j,
\label{eq:transform_fermion}
\eea
in which $S_T^z$ ($\in\mathbb{Z}$) is the total spin of the system.
In the transformation rules under $T$ and $R(\hat{y},\pi)$, the staggered sign $(-)^j$ comes from the factor of the string  $\Pi_{k<j}(2S_k^z)$. 
There is an indefinite sign $(-)^\#$ since the overall sign should be $\Pi_{k\leq j}(-)$,
which is not well-defined for a chain of infinite length (we need to consider an infinite chain so that linearization around Fermi points and low energy fields can be considered).

For a spin-1/2 XX chain, the system is converted into a non-interacting spinless fermion after the Jordan-Wigner transformation. 
The low energies physics is dominated by the two gapless points $\pm k_F$, in which $k_F=\pi/(2a)$ where $a$ is the lattice constant.  
Introducing the left and right movers $c_L^\dagger,c_R^\dagger$, the bosonized fields $\theta,\varphi$ can be related to the fermion fields as \cite{}
\bea
\varphi(x)&=&\varphi_0-(N_L+N_R)\sqrt{\pi}x/ L-i \sum_{q\neq 0} |\frac{1}{2qL}|^{1/2}\text{sgn}(q) e^{iqx} (b_q^\dagger+b_{-q}),\nn\\
\theta(x)&=&\theta_0+(N_L-N_R)\sqrt{\pi}x/L+i \sum_{q\neq 0} |\frac{1}{2qL}|^{1/2}e^{iqx} (b_q^\dagger-b_{-q}),
\label{eq:bosonization_fields}
\eea
in which $L$ is the system size; $N_r$ ($r=L,R$) is the (normal ordered) total fermion number in the $r$-branch
satisfying $S_T^z=N_L+N_R$; and
\bea
b_q^\dagger=|\frac{2\pi}{qL}|^{1/2} [\Theta(q)\rho_R^\dagger(q)+\Theta(-q)\rho_L^\dagger(q)],
\eea
where $\Theta$ is the step function and
\bea
\rho_r(q)=\sum_{k}c_{r,k+q}^\dagger c_{r,k}, ~r=L,R.
\eea
In Eq. (\ref{eq:bosonization_fields}),
the constant fields $\varphi_0$ and $\theta_0$ are the canonical conjugate fields of $N_L-N_R$ and $N_L+N_R$, respectively,
satisfying 
\bea
[N_L-N_R,\varphi_0]=[N_L+N_R, \theta_0]=\frac{i}{\sqrt{\pi}}.
\label{eq:commutation_0}
\eea
In fact, $\varphi_0$ represents the center of mass position of all the fermion, and $\theta_0$ represents the center of mass phase of the fermions. 

The translation operators are generated by the total momentum, i.e., $k_F(N_L-N_R)$.
It is clear that the translation operator does not change the components in $\theta(x),\varphi(x)$ except $\theta_0,\varphi_0$.
Hence $T_a=e^{ik_Fa(N_L-N_R)}=e^{i(N_L-N_R)\pi/2}$.
Using Eq. (\ref{eq:commutation_0}), we obtain
\bea
T_a \varphi_0 T_{-a}=\varphi_0+\frac{\sqrt{\pi}}{2}.
\label{eq:Ta_phi_0}
\eea
To derive how $T_a$ acts on $\theta_0$,
we need the following bosonization formula for $c^\dagger(x)$,
\begin{flalign}
c^\dagger(x)=\eta^\dagger_Re^{ik_Fx} e^{i\sqrt{\pi}[\theta(x)-\varphi(x)]}+\eta^\dagger_Le^{-ik_Fx} e^{i\sqrt{\pi}[\theta(x)+\varphi(x)]},
\label{eq:bosonize_fermion}
\end{flalign}
in which $\eta_R$ and $\eta_L$ are the Klein factors for the right and left movers.
Since $e^{\pm ik_Fx}$ differs from $e^{\pm ik_F(x+a)}$ by a factor of $\pm i$, we must have
\bea
T_a e^{i\sqrt{\pi}[\theta(x)-\varphi(x)]}T_{-a}&=& ie^{i\sqrt{\pi}[\theta(x)-\varphi(x)]},\nn\\
T_a e^{i\sqrt{\pi}[\theta(x)+\varphi(x)]}T_{-a}&=& -ie^{i\sqrt{\pi}[\theta(x)+\varphi(x)]}.
\eea
On the other hand, Eq. (\ref{eq:Ta_phi_0}) implies $T_a e^{i\sqrt{\pi}\varphi(x)}T_{-a}=ie^{i\sqrt{\pi}\varphi(x)}$, hence
$T_a e^{i\sqrt{\pi}\theta(x)}T_{-a}=- e^{i\sqrt{\pi}\theta(x)}$,
i.e.,
\bea
T_a\theta_0T_{-a} &=&\theta_0+\sqrt{\pi}.
\eea
Thus we obtain
\bea
T_a\theta(x)T_{-a}&=&\theta(x)+\sqrt{\pi},\nn\\
T_a\varphi(x)T_{-a}&=&\varphi(x)+\frac{\sqrt{\pi}}{2},\nn\\
T_a\eta_r^\dagger T_{-a}&=&\eta^\dagger_r,
\label{eq:transform_Ta}
\eea
in which $r=L,R$.

Define $\theta^\prime(x)$ ($\varphi^\prime(x)$) to be the components of $\theta(x)$ ($\varphi(x)$) excluding $\theta_0$ ($\varphi_0$). 
Since 
\bea
I\rho_r(q)I^{-1}&=&\rho_{-r}(-q)\nn\\
IN_rI^{-1}&=&N_{-r}
\eea
where $-L=R$ and $-R=L$.
For the inversion operation, it can be seen that 
\bea
I \theta^\prime (x)I^{-1}&=&\theta^\prime(-x)\nn\\
I \varphi^\prime (x)I^{-1}&=&-\varphi^\prime(-x).
\label{eq:I_prime}
\eea
To derive the transformations of $\theta_0,\varphi_0$,
we use Eq. (\ref{eq:bosonize_fermion}) to rewrite the second equation in Eq. (\ref{eq:transform_fermion}) and obtain
\begin{flalign}
&e^{ik_Fx}I\eta^\dagger_R e^{i\sqrt{\pi}[\theta(x)-\varphi(x)]}I^{-1}+e^{-ik_Fx}I \eta^\dagger_L e^{i\sqrt{\pi}[\theta(x)+\varphi(x)]}I^{-1}\nn\\
&=(\eta^\dagger_Re^{-ik_Fx} e^{i\sqrt{\pi}[\theta(-x)-\varphi(-x)]}+\eta^\dagger_Le^{ik_Fx} e^{i\sqrt{\pi}[\theta(-x)+\varphi(-x)]})e^{i\pi (N_L+N_R)},
\label{eq:Inversion_c}
\end{flalign}
which leads to
\bea
I\theta_0I^{-1}&=&\theta_0,\nn\\
I\varphi_0I^{-1}&=&-\varphi_0+\sqrt{\pi}(N_L+N_R),\nn\\
I\eta^\dagger_L I^{-1}&=&\eta^\dagger_R,\nn\\
I\eta^\dagger_R I^{-1}&=&\eta^\dagger_L.
\label{eq:derive_I}
\eea
Notice that in Eq. (\ref{eq:derive_I}), the fact that whether $e^{i\pi (N_L+N_R)}$ or $e^{-i\pi (N_L+N_R)}$ is used in Eq. (\ref{eq:Inversion_c}) does not matter.
Combining Eq. (\ref{eq:I_prime}) with Eq. (\ref{eq:derive_I}), we obtain
\bea
I\theta(x)I^{-1}&=&\theta(x),\nn\\
I\varphi(x)I^{-1}&=&-\varphi(x)+\sqrt{\pi}(N_L+N_R),\nn\\
I\eta^\dagger_r I^{-1}&=&\eta^\dagger_{-r}.
\eea

For time reversal, using 
\bea
T\rho_r(q)T^{-1}&=&-\rho_{-r}(-q),\nn\\
TN_rT^{-1}&=&-N_{-r},
\eea
we have
\bea
T \theta^\prime (x)T^{-1}&=&\theta^\prime(x)\nn\\
T \varphi^\prime (x)T^{-1}&=&-\varphi^\prime(x).
\label{eq:T_prime}
\eea
To derive the transformations of $\theta_0,\varphi_0$,
we use Eq. (\ref{eq:bosonize_fermion}) to rewrite the third equation in Eq. (\ref{eq:transform_fermion}) and obtain
\begin{flalign}
&e^{-ik_Fx}T\eta^\dagger_R e^{i\sqrt{\pi}[\theta(x)-\varphi(x)]}T^{-1}+e^{ik_Fx}T \eta^\dagger_L e^{i\sqrt{\pi}[\theta(x)+\varphi(x)]}T^{-1}\nn\\
&=(-)^\#[\eta_Re^{ik_Fx} e^{-i\sqrt{\pi}[\theta(x)-\varphi(x)]}+\eta_Le^{-ik_Fx} e^{-i\sqrt{\pi}[\theta(x)+\varphi(x)]}],
\label{eq:Inversion_c}
\end{flalign}
leading to
\bea
T\theta_0T^{-1}&=&\theta_0+\#\sqrt{\pi},\nn\\
T\varphi_0T^{-1}&=&-\varphi_0,\nn\\
T\eta^\dagger_RT^{-1}&=&\eta_L,\nn\\
T\eta^\dagger_LT^{-1}&=&\eta_R.
\label{eq:T_0}
\eea
The indefinite coefficient $\#$ can be determined  to be $1$ by imposing the condition $T\vec{S}_iT^{-1}=-\vec{S}_i$ from the bosonization formulas for the spin operators in Eq. (\ref{eq:abelian_bosonize}).
Combining Eq. (\ref{eq:T_prime}) with Eq. (\ref{eq:T_0}), we have
\bea
T\theta(x)T^{-1}&=&\theta(x)+\sqrt{\pi},\nn\\
T\varphi(x)T^{-1}&=&-\varphi(x),\nn\\
T\eta^\dagger_rT^{-1}&=&\eta_{-r}.
\eea

The generator for the U(1) transformation $R(\hat{z},\beta)$ is $N_L+N_R$.
Using the commutation relation in Eq. (\ref{eq:commutation_0}), it is straightforward to derive that 
\bea
R(\hat{z},\beta) \theta(x)[R(\hat{z},\beta)]^{-1}&=&\theta(x)+\frac{\beta}{\sqrt{\pi}},\nn\\
R(\hat{z},\beta) \varphi(x)[R(\hat{z},\beta)]^{-1}&=&\varphi(x),\nn\\
R(\hat{z},\beta) \eta_r^\dagger [R(\hat{z},\beta)]^{-1}&=&\eta_r^\dagger.
\eea

For $R(\hat{y},\pi)$, using
\bea
R(\hat{y},\pi)\rho_r(q)[R(\hat{y},\pi)]^{-1}&=&-\rho_r(q),\nn\\
R(\hat{y},\pi)N_r[R(\hat{y},\pi)]^{-1}&=&-N_r(q),
\eea
we obtain
\bea
R(\hat{y},\pi) \theta^\prime (x)[R(\hat{y},\pi)]^{-1}&=&-\theta^\prime(x)\nn\\
R(\hat{y},\pi) \varphi^\prime (x)[R(\hat{y},\pi)]^{-1}&=&-\varphi^\prime(x).
\label{eq:Ry_prime}
\eea
To derive the transformations of $\theta_0,\varphi_0$,
we use Eq. (\ref{eq:bosonize_fermion}) to rewrite the fifth equation in Eq. (\ref{eq:transform_fermion}) and obtain
\begin{flalign}
&e^{ik_Fx} R(\hat{y},\pi) \eta^\dagger_R e^{i\sqrt{\pi}[\theta(x)-\varphi(x)]}[R(\hat{y},\pi)]^{-1}+e^{-ik_Fx}R(\hat{y},\pi) \eta^\dagger_L e^{i\sqrt{\pi}[\theta(x)+\varphi(x)]}[R(\hat{y},\pi)]^{-1}=\nn\\
&(-)^\#[\eta_Re^{ik_Fx} e^{-i\sqrt{\pi}[\theta(x)-\varphi(x)]}+\eta_Le^{-ik_Fx} e^{-i\sqrt{\pi}[\theta(x)+\varphi(x)]}],
\label{eq:Ry_c}
\end{flalign}
leading to
\bea
R(\hat{y},\pi)\theta_0[R(\hat{y},\pi)]^{-1}&=&-\theta_0+\#\sqrt{\pi},\nn\\
R(\hat{y},\pi)\varphi_0[R(\hat{y},\pi)]^{-1}&=&-\varphi_0,\nn\\
R(\hat{y},\pi)\eta^\dagger_R[R(\hat{y},\pi)]^{-1}&=&\eta_R,\nn\\
R(\hat{y},\pi)\eta^\dagger_L[R(\hat{y},\pi)]^{-1}&=&\eta_L,
\label{eq:Ry_0}
\eea
where the indefinite sign can be determined in a similar way as the time reversal case to be $\sqrt{\pi}$.
Combining Eq. (\ref{eq:Ry_prime}) with Eq. (\ref{eq:Ry_0}), we have
\bea
R(\hat{y},\pi)\theta(x)[R(\hat{y},\pi)]^{-1}&=&-\theta(x)+\sqrt{\pi},\nn\\
R(\hat{y},\pi)\varphi(x)[R(\hat{y},\pi)]^{-1}&=&-\varphi(x),\nn\\
R(\hat{y},\pi)\eta^\dagger_r[R(\hat{y},\pi)]^{-1}&=&\eta_r.
\eea

In summary, we are able to derive the transformation properties in Eq. (\ref{eq:transformation_theta_phi}).
We note that the transformation rules of $\vec{S}_i$ in Eq. (\ref{eq:transform_spin}) can be obtained, by particularly noticing that rigorously,
the expression of $S^z_j$ is
\begin{flalign}
&S^{ z}(x)= -\frac{1}{\sqrt{\pi}} \nabla \varphi(x) + \text{const.} \frac{1}{a}(-)^n[\eta_R^\dagger \eta_L e^{i2\sqrt{\pi} \varphi(x)}+\eta_L^\dagger \eta_Re^{-i2\sqrt{\pi} \varphi(x)}].
\end{flalign}

\section{Determinations of magnetic ordering patterns}

In this appendix, we determine the patterns of magnetic orders using the unbroken symmetry groups,
by requiring that the spin expectation values be invariant under the unbroken symmetries.  

\subsection{Spiral order in Kitaev-Heisenberg chain}
\label{app:determine_orde1}

The unbroken symmetry group $H^{\text{(I)}}$ for the spiral order in Kitaev-Heisenberg chain in the $U_4$ frame is given in Eq. (\ref{eq:HI}),
containing four symmetry elements. 
By considering a four-site unit cell, it is enough to consider the four generators $T_aI$, $T_{2a}$, $[R(\hat{z},-\frac{\pi}{2})T_a]^2T$, and  $R(\hat{z},-\frac{\pi}{2})T_a\cdot R(\hat{y},\pi)T_aI$ (all modulo $T_{4a}$) for the group $(\mathbb{Z}_2\times \mathbb{Z}_2)\ltimes D^{(\text{I})}_2$ in Eq. (\ref{eq:short_exact_KH_spiral}). 
It can be verified that under the four generators, the transformations of the spin expectation values $\langle \vec{S}_i\rangle=(x_i,y_i,z_i)^T$ are 
\bea
T_{a}I&:&
\left(\begin{array}{c}
x_i\\
y_i\\
z_i
\end{array}
\right)\rightarrow
\left(\begin{array}{c}
x_{1-i}\\
y_{1-i}\\
z_{1-i}
\end{array}
\right),\nn\\
T_{2a}&:&
\left(\begin{array}{c}
x_i\\
y_i\\
z_i
\end{array}
\right)\rightarrow
\left(\begin{array}{c}
x_{i+2}\\
y_{i+2}\\
z_{i+2}
\end{array}
\right),\nn\\
(R(\hat{z},-\frac{\pi}{2})T_a)^2T&:&
\left(\begin{array}{c}
x_i\\
y_i\\
z_i
\end{array}
\right)\rightarrow
\left(\begin{array}{c}
x_{i+2}\\
y_{i+2}\\
-z_{i+2}
\end{array}
\right),\nn\\
R(\hat{z},-\frac{\pi}{2})T_a R(\hat{y},\pi)T_aI&:&
\left(\begin{array}{c}
x_i\\
y_i\\
z_i
\end{array}
\right)\rightarrow
\left(\begin{array}{c}
y_{2-i}\\
x_{2-i}\\
-z_{2-i}
\end{array}
\right).
\label{eq:unbroken_act1}
\eea

Invariance under $(R(\hat{z},-\frac{\pi}{2})T_a)^2T$ requires
\begin{flalign}
&x_1=x_3,~y_1=y_3,~z_1=-z_3\nn\\
&x_2=x_4,~y_2=y_4,~z_2=-z_4.
\label{eq:inv_3}
\end{flalign}
Invariance under $R(\hat{z},-\frac{\pi}{2})T_a R(\hat{y},\pi)T_aI$ requires
\begin{flalign}
&x_1=y_1,~z_1=0\nn\\
&x_3=y_3,~z_3=0\nn\\
&x_2=y_4,~y_2=x_4,~z_2=-z_4.
\label{eq:inv_4}
\end{flalign}
Combined with invariance under $T_aI$ and $T_{2a}$, we obtain
\begin{flalign}
&z_1=z_2=z_3=z_4=0\nn\\
&x_1=x_2=x_3=x_4=y_1=y_2=y_3=y_4.
\end{flalign}

\subsection{LLRR order in Kitaev-Heisenberg chain}
\label{app:determine_orde2}

The unbroken symmetry group $H^{\text{(II)}}$ for the spiral LLRR in Kitaev-Heisenberg chain in the $U_4$ frame is given in Eq. (\ref{eq:HII}),
generated by four elements $T_aI$, $T_{2a}$, $[R(\hat{z},-\frac{\pi}{2})T_a]^2T$, and $R(\hat{y},\pi)T_aI$.
The actions of $T_aI$, $T_{2a}$, and $[R(\hat{z},-\frac{\pi}{2})T_a]^2T$ on spin expectation values have been given in Eq. (\ref{eq:unbroken_act1}),
and the action of $R(\hat{y},\pi)T_aI$ is 
\bea
R(\hat{y},\pi)T_aI&:&
\left(\begin{array}{c}
x_i\\
y_i\\
z_i
\end{array}
\right)\rightarrow
\left(\begin{array}{c}
-x_{1-i}\\
y_{1-i}\\
-z_{1-i}
\end{array}
\right). 
\eea
Invariance under $R(\hat{y},\pi)T_aI$ requires
\begin{flalign}
&x_1=-x_4,~y_1=y_4,~z_1=-z_4\nn\\
&x_2=-x_3,~y_2=y_3,~z_2=-z_3.
\label{eq:unbroken_act2}
\end{flalign}
Combining the four generators, the magnetic ordering invariant under $H^{\text{(II)}}$ is given by
\begin{flalign}
&x_1=x_2=x_3=x_4=0\nn\\
&y_1=y_2=y_3=y_4\nn\\
&z_1=z_2=z_3=z_4=0.
\end{flalign}
This is the pattern given in Eq. (\ref{eq:order_S_prime_general_II}).

\subsection{Distorted spiral order in Kitaev-Heisenberg-Gamma chain}
\label{app:determine_orde3}

In the Kitaev-Heisenberg-Gamma chain, the unbroken symmetry group $H_1^{\text{(I)}}$ for the distorted spiral order is generated by $[R(\hat{z},-\frac{\pi}{2})T_a]^2T$ and  $R(\hat{z},-\frac{\pi}{2})T_a\cdot R(\hat{y},\pi)T_aI$,
hence only invariances in Eq. (\ref{eq:inv_3},\ref{eq:inv_3}) are required, leading to
\begin{flalign}
&x_1=y_1=x_3=y_3,~z_1=z_3=0\nn\\
&x_2=y_2=x_4=y_4,~z_2=-z_4.
\end{flalign}
This is the pattern given in Eq. (\ref{eq:Spin_align_d_sp_4rot}).

\subsection{Distorted LLRR order in Kitaev-Heisenberg-Gamma chain}
\label{app:determine_orde4}

In the Kitaev-Heisenberg-Gamma chain, the unbroken symmetry group $H_1^{\text{(II)}}$ for the distorted LLRR order is generated by $[R(\hat{z},-\frac{\pi}{2})T_a]^2T$ and  $R(\hat{y},\pi)T_aI$,
hence only invariances in Eq. (\ref{eq:unbroken_act1},\ref{eq:unbroken_act2}) are required, leading to
\begin{flalign}
&x_1=-x_2=x_3=-x_4\nn\\
&y_1=y_2=y_3=y_4\nn\\
&z_1=z_2=-z_3=-z_4.
\end{flalign}
This is the pattern given in Eq. (\ref{eq:Spin_align_d_sp_4rot_II}). 

\subsection{FM I order in Kitaev-Heisenberg-Gamma chain}
\label{app:determine_FM_I}

In the Kitaev-Heisenberg-Gamma chain, the unbroken symmetry group $H_2^{\text{(I)}}$ for the FM I order is generated by 
$T_aI$ and $R(\hat{n}_1,\pi)T_a$ as discussed in Eq. (\ref{eq:H_2^II}).
The spin expectation values transform under these two operations according to
\bea
T_{a}I&:&
\left(\begin{array}{c}
x_i\\
y_i\\
z_i
\end{array}
\right)\rightarrow
\left(\begin{array}{c}
x_{1-i}\\
y_{1-i}\\
z_{1-i}
\end{array}
\right),\nn\\
R(\hat{n}_1,\pi)T_a&:&
\left(\begin{array}{c}
x_i\\
y_i\\
z_i
\end{array}
\right)\rightarrow
\left(\begin{array}{c}
-y_{i+1}\\
-x_{i+1}\\
-z_{i+1}
\end{array}
\right).
\eea
By considering a two-site unit cell, the invariance under $T_aI$ requires
\bea
x_1=x_2,~y_1=y_2,~z_1=z_2,
\label{eq:FMI_invariance_1}
\eea
and the invariance under $R(\hat{n}_1,\pi)T_a$ requires
\bea
x_1=-y_2,~y_1=-x_2,~z_1=-z_2.
\label{eq:FMI_invariance_2}
\eea
Combining Eq. (\ref{eq:FMI_invariance_1}) and Eq. (\ref{eq:FMI_invariance_2}), we obtain the pattern of the magnetic ordering in Eq. (\ref{eq:spin_FM_general_I}). 

\subsection{FM II order in Kitaev-Heisenberg-Gamma chain}
\label{app:determine_FM_II}

In the Kitaev-Heisenberg-Gamma chain, the unbroken symmetry group $H_2^{\text{(II)}}$ for the FM I order is generated by 
$T_aI$ and $TR(\hat{n}_1,\pi)T_a$ as discussed in Eq. (\ref{eq:H_2^II_2}).
The spin expectation values transform under $TR(\hat{n}_1,\pi)T_a$ according to 
\bea
TR(\hat{n}_1,\pi)T_a&:&
\left(\begin{array}{c}
x_i\\
y_i\\
z_i
\end{array}
\right)\rightarrow
\left(\begin{array}{c}
y_{i+1}\\
x_{i+1}\\
z_{i+1}
\end{array}
\right).
\eea
The invariance under $TR(\hat{n}_1,\pi)T_a$ requires
\bea
x_1=y_2,~y_1=x_2,~z_1=z_2.
\label{eq:FMI_invariance_3}
\eea
Combining Eq. (\ref{eq:FMI_invariance_1}) with Eq. (\ref{eq:FMI_invariance_3}), 
we obtain the pattern of the magnetic ordering in Eq. (\ref{eq:spin_FM_general}). 

\end{widetext}


\end{document}